\documentclass[aps,pre,reprint,showpacs,superscriptaddress]{revtex4-1}
\usepackage{graphicx}% Include figure files

\usepackage{dcolumn}% Align table columns on decimal point
\usepackage{bm}% bold math
\usepackage{color}
\usepackage{lipsum}
\usepackage{mathrsfs}
\usepackage{lineno}
\usepackage{subfig}
\usepackage{amsmath}
\usepackage{float}
\usepackage[colorlinks,urlcolor=blue,linkcolor=blue,citecolor=blue]{hyperref}
\usepackage{siunitx}
\usepackage{multirow}
\begin{document}
\captionsetup[figure]{justification=raggedright,singlelinecheck=false}
\graphicspath{{picture/}}  % Place your graphic files in the same directory as your main document
\DeclareGraphicsExtensions{.pdf, .jpg, .tif, .png}

% For including Chinese characters, try one of the following:
%\begin{CJK*}{GB}{gbsn}
%\begin{CJK*}{GB}{}
% To insert Chinese characters, use any of the Chinese character soft keyboards, such as 
% Title of paper
%\linenumbers

\title{Non-Equilibrium and Self-Organization Evolution in Hot-Spot Ignition Processes}
% Ignition of Non-equilibrium Hot Spot Models and Emergent Self-Organization Phenomena
% The Impact of Non-equilibrium Hot Spot Models on Ignition Margin and Emergent Self-organization Phenomena
% 非平衡模型产生的点火裕度以及非平衡自组织现象涌现

%\email, \thanks, \homepage, \altaffiliation all apply to the current
% Please use the appkropriate macro for each type of information
%\affiliation command applies to all authors since the last
%\affiliation command. The \affiliation command should follow the
%\affiliation can be followed by \email, \homepage, \thanks as well.
%\thanks{}
\author{X.-Y. Fu}
% \affiliation{Collaborative Innovation Center of IFSA (CICIFSA), Shanghai Jiao Tong University, Shanghai 200240, People’s Republic of China}

\author{Z.-Y. Guo}
% \affiliation{Collaborative Innovation Center of IFSA (CICIFSA), Shanghai Jiao Tong University, Shanghai 200240, People’s Republic of China}
\author{Q.-H. Wang}
% \affiliation{Collaborative Innovation Center of IFSA (CICIFSA), Shanghai Jiao Tong University, Shanghai 200240, People’s Republic of China}
\author{R.-C. Wang}
% \affiliation{Collaborative Innovation Center of IFSA (CICIFSA), Shanghai Jiao Tong University, Shanghai 200240, People’s Republic of China}
\author{D. Wu}
\email{dwu.phys@sjtu.edu.cn}
\affiliation{Key Laboratory for Laser Plasmas and Department of Physics and Astronomy, Collaborative Innovation Center of IFSA (CICIFSA), Shanghai Jiao Tong University, Shanghai 200240, People’s Republic of China}
\affiliation{Zhiyuan College, Shanghai Jiao Tong University, Shanghai 200240, People’s Republic of China}
\author{J. Zhang}
\email{jzhang@iphy.ac.cn}
\affiliation{Key Laboratory for Laser Plasmas and Department of Physics and Astronomy, Collaborative Innovation Center of IFSA (CICIFSA), Shanghai Jiao Tong University, Shanghai 200240, People’s Republic of China}
\affiliation{Zhiyuan College, Shanghai Jiao Tong University, Shanghai 200240, People’s Republic of China}
\affiliation{Institute of Physics, Chinese Academy of Sciences, Beijing 100190, People’s Republic of China}

\date{\today}
\begin{abstract}
% In the double-cone ignition regime, the compressed fuel is believed to be at a state far beyond the description of classical plasmas, in which the quantum degeneracy plays important roles. 
Due to disparate formation mechanisms, as for central hot-spot ignition and fast ignition, the initial temperatures of electron and ions usually differs from each other in the hot spot. 
Considering the percipient dependence of fusion cross-section and energy losses on temperature, this difference manifests the inadequacy of the equilibrium theoretical model in accurately depicting the ignition condition and evolution of the hot-spot.
In this work, we studied a non-equilibrium model and extended this model to both isobaric and isochoric scenarios, characterized by varying hot-spot densities, temperatures and expansion velocities.
In both cases, a spontaneous self-organization evolution was observed, manifesting as the bifurcation of ion and electron temperatures. Notably, the ion temperature is particularly prominent during the ignition process.
This inevitability can be traced to the preponderant deposition rates of alpha-particles into D-T ions and the decreasing rate of energy exchange between electrons and D-T ions at elevated temperatures. The inherent structure, characterized by higher ion temperature and lower electron temperature during ignition, directly contributes to the augmentation of D-T reactions and mitigates energy losses through electron conduction and bremsstrahlung, thereby naturally facilitating nuclear fusions.
% 非平衡条件下isobaric and isochoric的共同特征和差异特征
% 重点强调，无论何种条件，都出现了自组织现象，分析这种现象出现的必然性，并强调这一发现的重要意义。

\end{abstract}


\maketitle

\section{Introduction}

In practical researches, there exist two distinct implosion design strategies, i.e., central hot-spot ignition and fast ignition, characterized by unique processes for the formation of hot-spots \cite{atzeni_inertial_2013}.
In the case of central hot-spot ignition, which is powered by X-rays or lasers, the D-T gas, along with the surrounding high density D-T fuel, is compressed inward, leading to the formation of a high-temperature hot spot at the core of the fuel. Conventionally, the hot spot achieved through this design maintains nearly constant pressure compared to the surrounding cold fuels \cite{clavin_quasi-isobaric_2017}. During the implosion process, the majority of the shock wave energy is deposited into ions, resulting in elevated ion temperatures, as experimentally verified \cite{rygg_electron-ion_2009}. 
In contrast, the fast ignition scheme separates the compression and hot-spot formation processes \cite{tabak_fast_2006, atzeni_inertial_2013, ghasemi_analytical_2014}.
During the compression, the density remains uniform, necessitating the employment of an isochoric design \cite{xu_formation_2023, clark_self-similar_2007-1, farahbod_improvement_2014}.
After the compression process, a beam of fast electrons is injected and deposited, resulting in the formation of a localized hot-spot with electron temperature significantly higher than that of the ions \cite {ghasemi_analytical_2014}.

The central hot-spot design has remained the prevailing strategy and has attained several milestones in recent years, e.g. burning plasma state \cite{zylstra_burning_2022}, 
ignition \cite{acree_lawson_2022} and ``scientific breakeven'' \cite{acree_achievement_2024}.
Despite the achievement achieved, there are still unresolved facets of new physics within the burning plasmas and ignition processes, as detailed in \cite{zylstra_burning_2022, hurricane_fuel_2014, lindl_review_2014}. These include kinetic effects and the energy transfer mechanisms of $\alpha$-particles during the self-burning processes, as evidenced in \cite{hartouni_evidence_2023}. 

Concerning the unresolved ignition queries, previous theoretical efforts have primarily resorted to the simplified equilibrium model \cite{chang_generalized_2010, doppner_demonstration_2015, gopalaswamy_demonstration_2024, churnetski_three-dimensional_2024, atzeni_physics_2004}. 
For the sake of simplicity, it has become a common assumption to presume nearly identical ion and electron temperatures throughout the entire ignition process \cite{hurricane_physics_2023, daughton_infuence_2023}.
However, this approach seems to overlook the likelihood of non-equilibrium ion and electron temperatures.
We hypothesize that, at high temperatures, the equilibrium condition might be disrupted due to different heating mechanisms acting on ions and electrons, resulting in distinct evolution.

% 第四段，介绍一下Z. F. Fan的工作，进而引出我们的工作。
In Z. F. Fan's work, the ion-electron non-equilibrium ignition was theoretically studied through the introduction of a two-temperature mode \cite{fan_ignition_2016}. 
They consider separate thermal equilibrium of ions and electrons at temperatures $T_i$ and $T_e$ ($T_i \neq T_e$).
Consequently, these two components evolve independently with energy exchange adjusting between them due to ion-electron collisions.
According to their findings, an initially higher hot-spot ion temperature compared to electrons enhances nuclear reactions and reduces energy loss of electron conduction and bremsstrahlung, thereby facilitating ignition.
Building upon their model, we further develop an isochoric model for fast ignition, depict the ignition condition in different cases, and provide a detailed analysis of different heating stages during both hot-spot ignition and fast ignition processes. 
In the context of fast ignition, the disparity between ion and electron temperatures exerts a significant influence on the heating process due to extreme density, offering valuable insights for optimizing the hot-spot formation process.
Additionally, disparate deposition rates of alpha-particle heating to ions and electrons lead to the bifurcated evolution of ion and electron temperatures, ultimately and naturally benefiting ignition.

% 第五段，我们做了什么
The organization of this paper is as follows. In Section \ref{sec:Non-Equilibrium Model}, we introduce the equations that form the basis of our non-equilibrium model.
Subsequently, Section \ref{sec:Simulation results} offers an overview of the data sources employed and presents our simulation outcomes for both the isobaric and isochoric models.
Furthermore, Section \ref{sec:Full Analysis Different heating stages} delves into a theoretical analysis of the distinct temporal stages observed in the heating process.
Lastly, Section \ref{sec:Conclusion and discussions} concludes the paper by summarizing our key findings and discussing their implications.

%(Introducing the equilibrium model and pointing out its deficiency. In the previous study they also considered the non-equilibrium model, we say we follow the same routine, extend it to the isochoric model, and find some interesting phenomenon, e.g. Self-organized criticality, on the heating process.)

%(A little introduction of our model. How do we consider the isochoric model? Introduce our advantages and new limits on ignition conditions. Introduce the interesting phenomenon, e.g. Self-organized criticality. Future application.)

\section{Non-Equilibrium Model}
\label{sec:Non-Equilibrium Model}

The hot-spot, characterized by its radius $R_h$ and total density $\rho_h$, can be simplified as comprising two components: D-T ions and electrons. Assuming ion temperatures and pressures are denoted as $T_i,~P_i$ and electron temperatures and pressures as $T_e,~P_i$, we define the hot-spot equilibrium temperature as $T_h \equiv (T_i+T_e)/2$ and introduce a non-equilibrium factor, $f=T_i/T_h$.

During ignition, the alpha particles produced by D-T reactions, primarily concerned with the ion temperature $T_i$, deposit energy into the hot spot at a total power of \(W_\alpha=A_\alpha \rho_h^2 \langle \sigma v \rangle f_\alpha\), with \(A_\alpha = 8 \times 10^{40} ~{\rm erg/g^2}\).
Here, $\langle \sigma v \rangle$ is the cross-section of D-T fusion, expressed as a function of $T_i$ \cite{atzeni_inertial_2013}. The deposition rate of $\alpha$ in hot spots $f_{\alpha}$ has a form of 
\begin{equation}
    f_\alpha
= \begin{cases}({3}/{2}) \tau_\alpha-({4}/{5}) \tau_\alpha^2, & \tau_\alpha \leq 1 / 2 \\ 1-({1/4}){\tau_\alpha}^{-1}+({1}/{160}) {\tau_\alpha^{-3}}, & \tau_\alpha \geq 1 / 2\end{cases}, 
\end{equation}
where $\tau_\alpha\simeq 9 \times{\ln \Lambda } {\rho_{h} R_{\mathrm{h}}}/{T_{\mathrm{h}}^{3 / 2}}$. 
This energy deposition heats the ions and electrons in the hot-spot by fractions of $f_{\alpha i}=T_e/(32+T_e)$ and $f_{\alpha e}=1-f_{\alpha i}$, respectively, where $T_e$ is given in keV \cite{fraley_thermonuclear_1974}.
The energy losses in the system occur through electron thermal conduction with the surrounding main fuel, given by $W_e=3A_e T_e^{7/2}/R_h^2/\ln \Lambda$ with $A_e= 9.5 \times 10^{19} ~{\rm{erg~s^{-1}cm^{-1}keV^{-7/2}}}$, and through electron bremsstrahlung $W_r=A_r \rho_h^2 T_e^{1/2}$ with $A_r= 3.05 \times 10^{23}~{\rm erg~cm^{3} g^{-2} s^{-1} keV^{-1/2}}$. 
The Coulumb logarithm $\ln\Lambda$ can be expressed as $
  0.5\times\ln\left[1+({b_{\min}}/{b_{\max}})^2 \right]
$ \cite{wu_particle--cell_2018}
where $b_{\min}$ is the maximum between the classical impact parameter and the De Broglie wavelength, 
and $b_{\max}$ is the Debye length.
Additionally, the expansion of the ions and electrons within the hot spot, together with speed $u_h$, contributes to energy loss with powers of $W_{m,i}= 3 P_i u_h/R_h$ and $W_{m,e}= 3 P_e u_h/R_h$, respectively.
Furthermore, energy exchange between ions and electrons occurs through collisions, contributing to a power of $W_{ie} = A_{\omega,ei} \rho_h^2 \ln \Lambda (T_i - T_e )/T_e^{3/2}$, where $A_{\omega,ei}= 5.6 \times 10^{14}~{\rm kJ~cm^3 g^{-2} s^{-1} keV^{1/2}}$.

The temporal evolution for the temperatures of ion and electron energies, accounting for both the energy gains and losses in the hot spot, are expressed as follows \cite{fan_ignition_2016}:
\begin{align}
  C_{V,i} \rho_i \frac{\mathrm d T_i}{\mathrm d t}
&=
  W_\alpha f_{\alpha i} - W_{ie} - W_{m,i}, 
\label{eq:ion-temperature}
  \\
  C_{V,e} \rho_e \frac{\mathrm d T_e}{\mathrm d t}
&=
  W_\alpha f_{\alpha e} + W_{ie} - W_{m,e} -W_r -W_e, 
\label{eq:electron-temperature}
\end{align}
where $C_{V,i}$ and $C_{V,e}$ are the specific heat of ions and electrons respectively; $\rho_i$ and $\rho_e$ are the density of ions and electrons.

Energy spreading outside the hot spot, include electron bremsstrahlung power $W_r$ and electron thermal conduction $W_e$. The latter contributes to main fuel ablation, ultimately increasing the hot spot mass, which is expressed as \cite{hurricane_physics_2023,daughton_infuence_2023,spears_influence_2008}:
\begin{equation}
  \frac{\mathrm d \rho_h V_h}{\mathrm d t}
=
  \frac{[W_r + W_e + W_\alpha(1-f_\alpha)/f_\alpha]V_h }{C_{v,i} T_i + C_{v,e}T_e}.
\label{eq:ablation}
\end{equation}
Our primary interest lies in the density variation, therefore, we formulate the contribution of volume $V_h$ and rewrite Eq.\ \eqref{eq:ablation} as
\begin{equation}
  \frac{\mathrm d \rho_h}{\mathrm d t}
=
  \frac{W_r + W_e + W_\alpha(1-f_\alpha)/f_\alpha }{C_{v,i} T_i + C_{v,e}T_e}
  - \frac{ 4 \pi R_h^2 \rho_h u_h}{V_h}, 
\label{eq:rhoh}
\end{equation}
considering the hot-spot expansion rate given by
\begin{equation}
  \frac{\mathrm d R_h}{\mathrm d t} 
=
  u_h.
\label{eq:Rh}
\end{equation}

\begin{figure}[htbp]
    \centering
    \includegraphics[width=1.0\linewidth]{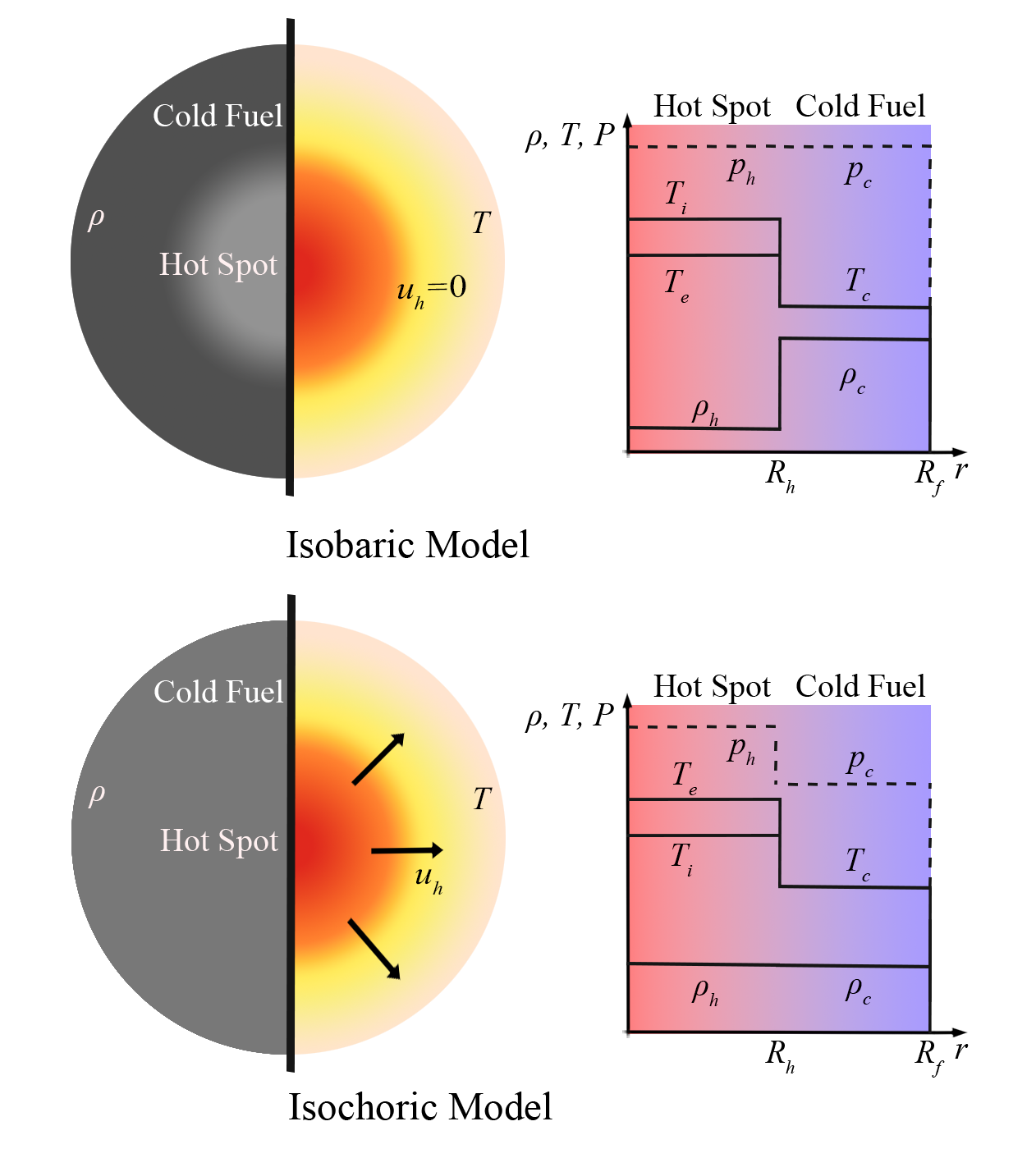}
    \caption{Schematic description of isobaric and isochoric hot-spot ignition models. }
    \label{fig:model disgram}
\end{figure}
In our research, significant attention is focused on the stagnation moment, which lies between the end of compression and the decomposition of the hot spot. This transient yet pivotal moment exhibits distinct behaviors in response to various ignition approaches. In the case of central hot-spot ignition, the hot spot follows the isobaric description, where its pressure approximates that of the surrounding cold fuel, resulting in minimal alterations to its radius \cite{daughton_infuence_2023, atzeni_inertial_nodate}. Conversely, the fast ignition scheme features a hot spot density that closely mirrors the cold fuel, utilizing the isochoric model \cite{xu_formation_2023}. Consequently, hot spots expand due to pressure differentials. The intricate relationship between these models is graphically represented in Fig.\ \ref{fig:model disgram}. The respective expansion rates for these models are mathematically described in \cite{atzeni_inertial_2013},
\begin{align}
  u_h
=
\begin{cases}
\sim0, & \text{Isobaric Model}\\
\sim\displaystyle{\left(\frac 3 4 \Gamma_B T_h \frac{\rho_h}{\rho_c}\right)^{1/2}}, & \text{Isochoric Model}
\end{cases}
\end{align}
where $\Gamma_B$ is related to the Boltzmann constant, for D-T fuel, $\displaystyle{\Gamma_B = 4k_B/(2m_e+5m_p) = 7.66\times 10^{14} ~\mathrm{erg/(g~keV)}}$; $\rho_h$ and $\rho_c$ are respectively the density of the hot spot and the main fuel.

By solving the aforementioned set of four differential equations, we obtain valuable insights into the ignition processes, which are elucidated in Section \ref{sec:Simulation results}.

\section{Numerical results}
\label{sec:Simulation results}

%等压等容的scan

%等压等容升温曲线

%f对升温速度的影响

%选取等容和等压各一个升温曲线，画不同时间阶段的功率图几个功率放一起，保留数据。

In the previous section, we have listed a set of differential equations for temporal evolution of physical quantities. In this sections, this set of equations are numerically solved and displayed. In Table \ref{tab:simulated configuration}, the initial parameters are displayed, and especially, the areal density is kept the same in both models for a consistent comparison between them. 

\begin{table}[hbtp]
\caption{
Numerical configurations of the initial state, including the radius, density and temperature of the hot spot, for both isobaric and isochoric models.
}
\begin{ruledtabular}
\begin{tabular}{rccc}
\textrm{Scheme}&
$R_h$ ($\si{\mu m}$)&
$\rho_h$ ($\si{g/cm^3}$)&
$T_h$ ($\si{keV})$\\
\colrule
Isobaric Model & 100 & 100 & 8 \\
Isochoric Model & 10 & 1000 & 8 \\
\end{tabular}
\label{tab:simulated configuration}
\end{ruledtabular}
\end{table}

To investigate the non-equilibrium dynamics comprehensively, we conduct numerical analyses with initial states of various conditions for both isochoric and isobaric models. Our study involves analyzing the temporal evolution of electron temperature and ion temperature. Additionally, we explore the distribution of initial values leading to successful ignition within the phase space (areal density and temperature) over a finite time. These analyses aim to delineate the ignition thresholds associated with distinct equilibrium factors and models.

\subsection{Isobaric Model}

\begin{figure}[htbp]
    \centering
    \includegraphics[width=0.95\linewidth]{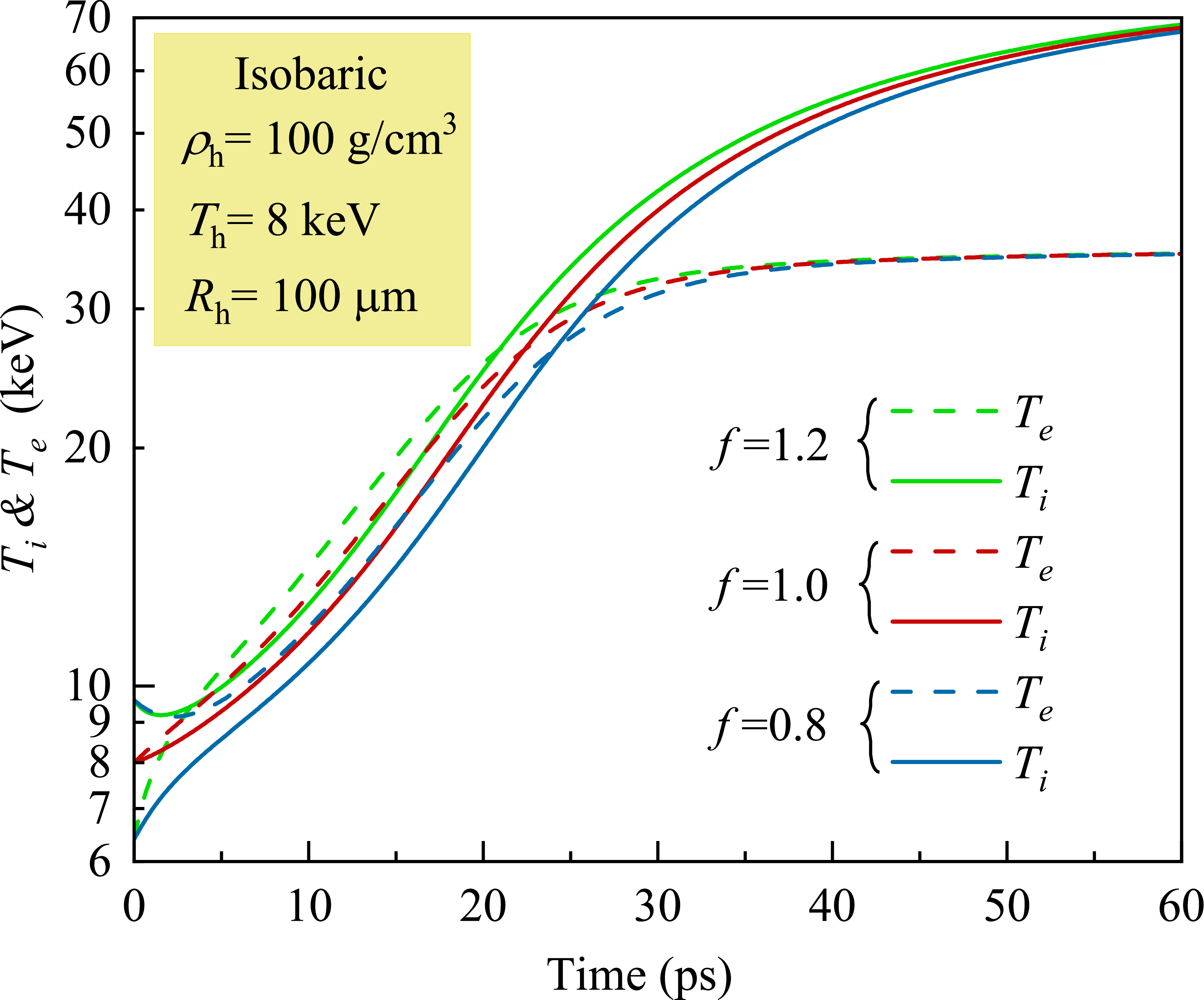}
    \caption{In the isobaric model, with areal density of $\rho_h R_h = 10\mathrm{g/cm^2}$, the ion and electron temperature rise as a function of time for different initial non-equilibrium factors $f=0.8$, $1$ and $1.2$. These two temperatures would finally become bifurcated and reach saturated values.}
    \label{fig:enter-label11}
\end{figure}

Figure\ \ref{fig:enter-label11} is the temperature evolution for the isobaric model. From this evolution curve, we can find that electrons and ions first undergo energy exchange with each other, similar to a relaxation process \cite{daligault_theory_2019}. Then the temperature of two species rise steadily until it reaches a certain value. During the entire process, a remarkable phenomenon occurs: after reaching a specific value, the temperatures of ions and electrons spontaneously diverge, ultimately reaching their saturated states. This heating process can be therefore broadly categorized into four distinct stages, and in the following we will elucidate each of them.

Under the influence of the non-equilibrium factor, we can find that the ignition is more prominent when the non-equilibrium factor is high, which means the initial temperature of ions is higher than that of electrons.

\subsection{Isochoric Model}
\label{sec:Isochoric Model}
\begin{figure}[htbp]
    \centering
    \includegraphics[width=0.95\linewidth]{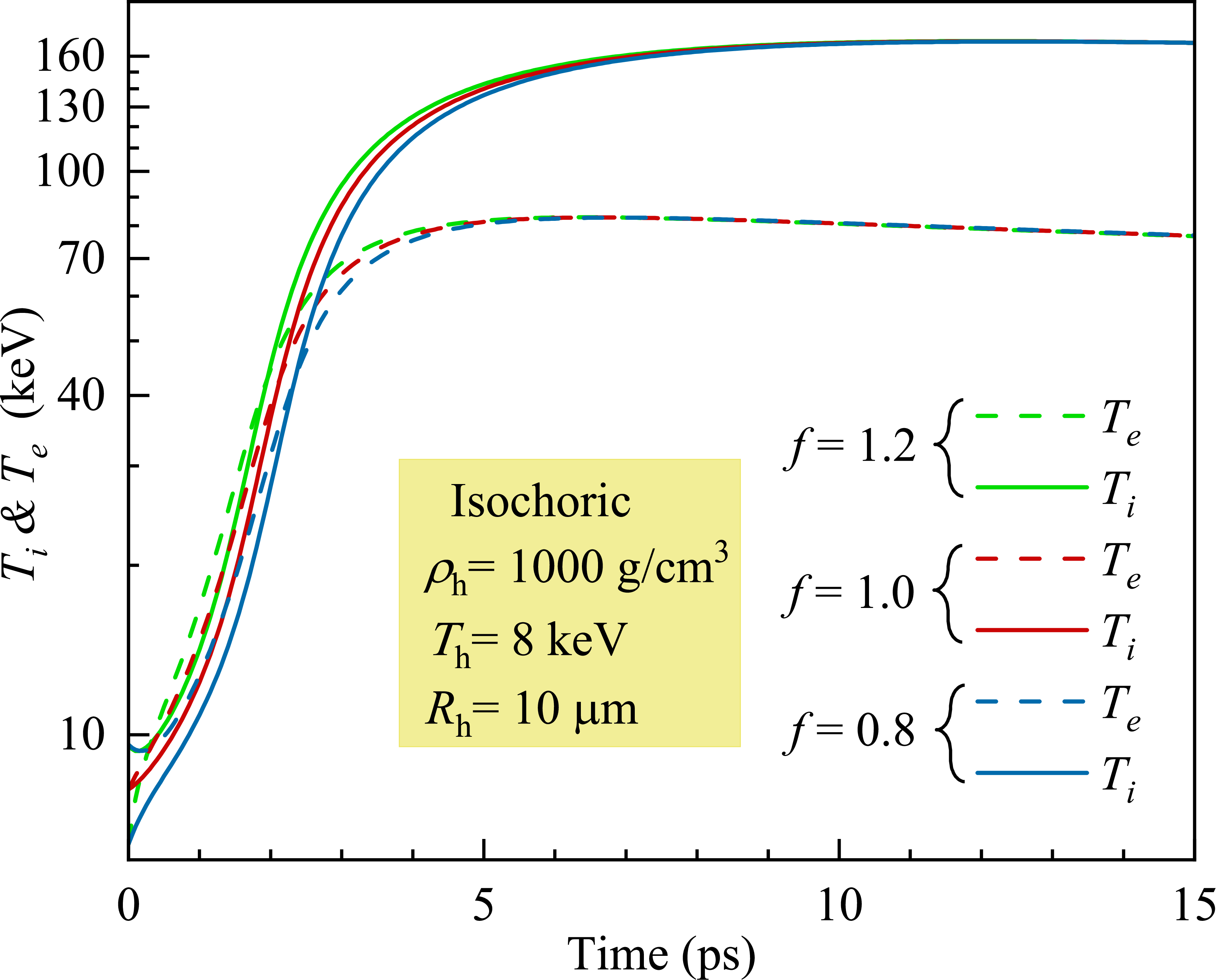}
    \caption{In the isochoric model with an areal density $\rho_h R_h = 10\ \mathrm{g/cm^2}$, the ion and electron temperature rise with time at different initial non-equilibrium factors $f=0.8$, $1$, and $1.2$. These two temperatures will become bifurcated and eventually decreased due to expansion.}
    \label{fig:enter-label2}
\end{figure}
Figure \ref{fig:enter-label2} reveals that the temporal evolution of temperatures for the isochoric model is quite similar with that of the isobaric model. The influence of the non-equilibrium factor remains consistent with that observed in the isobaric model. Notably, the temperature of ions plays a pivotal role in the heating process. However, there still exits two main differences. Firstly, if the time duration is significantly extended, the temperature of both ions and electrons will decrease as a result of the expansion of the hot-spot, and this trend is more significant in the isochoric model. Secondly, by comparing the first 5 ps in Fig.\ \ref{fig:enter-label11} and Fig.\ \ref{fig:enter-label2}, we observe significant differences in between the isobaric and isochoric models. Notably, in the isochoric model, DT hot-spot heats up considerably faster than the isobaric model, resulting in a much higher peak temperature. 

\subsection{Ignition Condition}

The ignition threshold curve establishes the boundaries within the initial phase space (areal density and temperature) that permit successful ignition within a finite time span. Specifically, an excessively low areal density is precluded as it hinders electron heat conduction at elevated temperatures, as discussed in \cite{temporal_ignition_2012}. Conversely, a high areal density enables the temperature to approach closely the critical value of 4.3 keV, which is essential for self-heating.

\begin{figure}[htbp]
    \centering
    \includegraphics[width=0.95\linewidth]{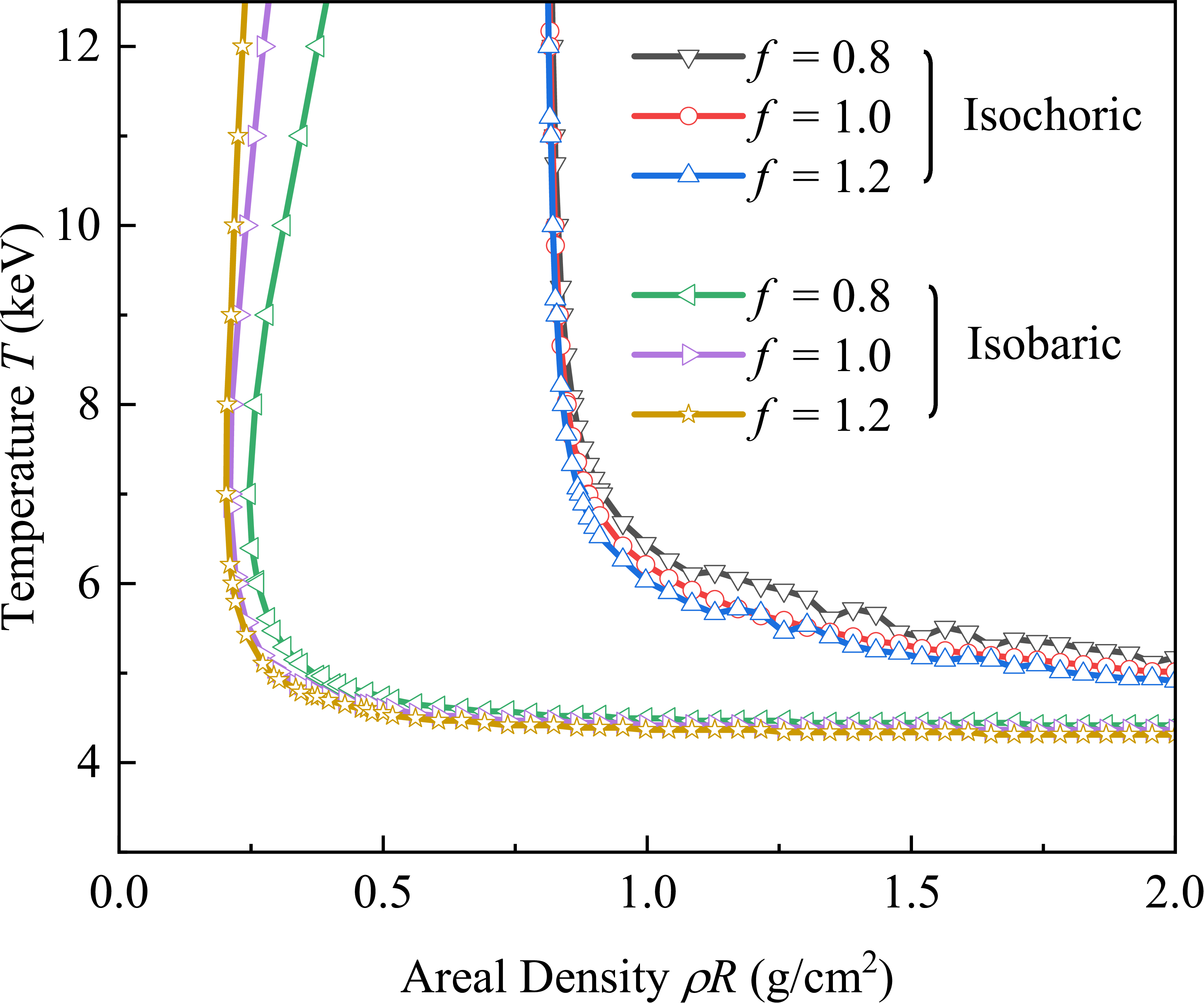}
    \caption{The ignition condition for both isobaric and isochoric models. When the initial condition of the hot spot is above the curve, the ignition would take place.}
    \label{fig:In this isobaric heating model, three different}
\end{figure}

In comparison, the isobaric model exhibits less stringent ignition requirements than the isochoric model. Regardless of the model chosen, an increase in the non-equilibrium factor plays a pivotal role in expanding the ignition area. This underscores the positive impact of enhancing the non-equilibrium factor, which effectively lowers ignition thresholds.

\section{Analysis of Ignition Evolution}
\label{sec:Full Analysis Different heating stages}
Our focus will primarily be on analyzing the isobaric model. As mentioned in Section\ \ref{sec:Isochoric Model}, the isochoric model exhibits similarities to the isobaric model in terms of its physical characteristics. However, a key distinction lies in the final process, which is influenced by the dominance of expansion power.

\begin{figure*}[htbp]
    \centering
    \includegraphics[width=0.95\linewidth]{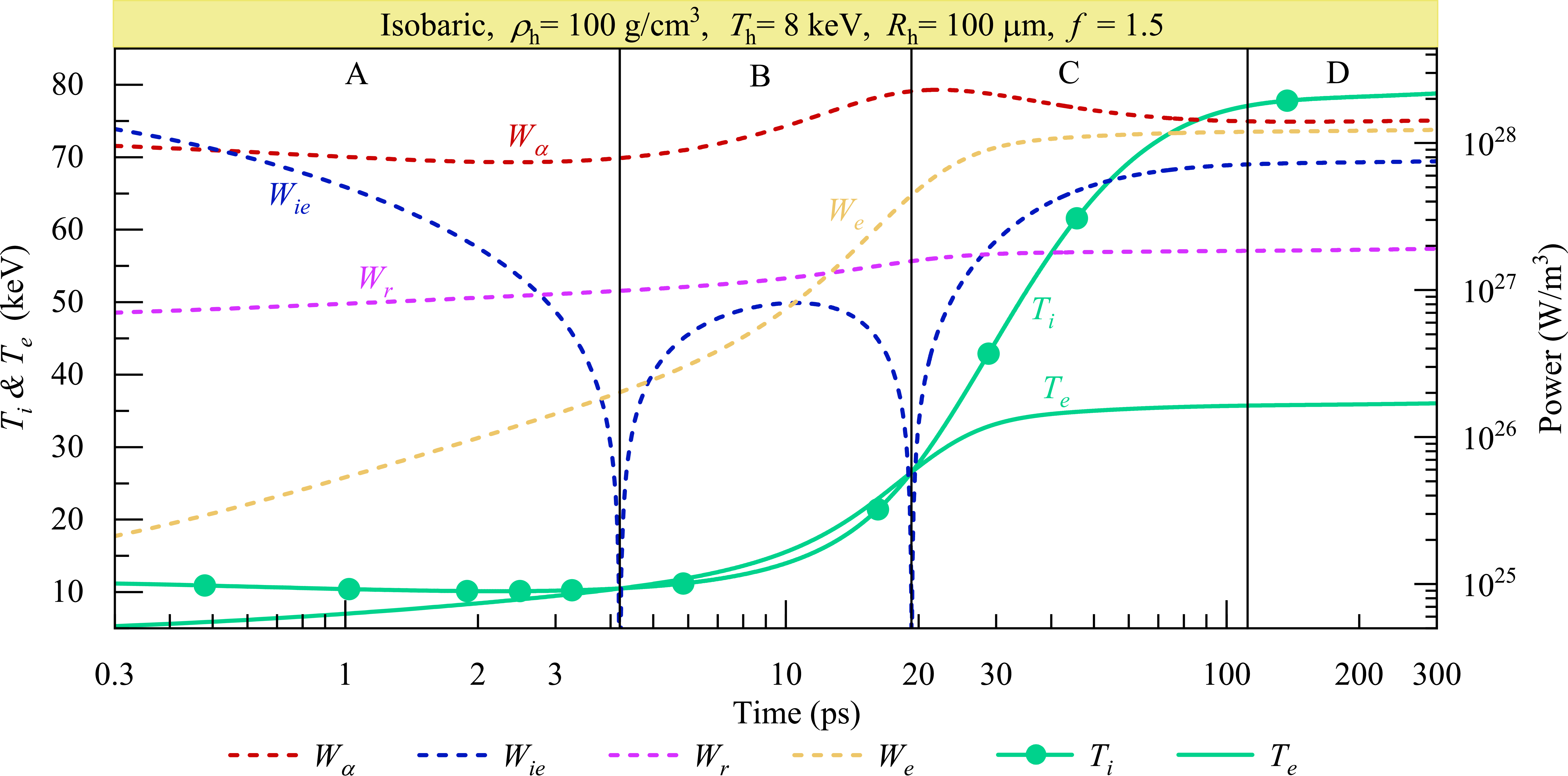}
    \caption{The ignition process for an isobaric model, with $T_i$, $T_e$, $W_{\alpha}$, $W_{ie}$, $W_{r}$ and $W_{e}$ evolving as a function of time.}
    \label{fig:stages}
\end{figure*}

In the ignition process, depicted in Fig.\ \ref{fig:stages}, intriguing phenomena emerge, indicating that the heating process can be segmented into four distinct stages: thermal equilibrium (Stage A), co-heating (Stage B), bifurcated heating (Stage C), and attaining saturated temperatures (Stage D). We aim to delve into each of these processes and provide a theoretical analysis. Fig.\ \ref{fig:stages} also illustrates the power per volume throughout the ignition process. Notably, the expansion work in the isobaric model is ignored. The exchange energy between ions and electrons, however, is influenced by both the temperature of the ions and electrons, necessitating the utilization of an absolute value for this particular energy component.

\subsection{Reaching a Dynamic Equilibrium}

At the beginning, the average temperature of ions and electrons does not increase significantly, but the temperature difference gradually decreases, as presented in the Stage A of Fig.\ \ref{fig:stages}.
We therefore subtract Eq.\ \eqref{eq:ion-temperature} and Eq.\ \eqref{eq:electron-temperature}, to derive the temperature differences between ions and electrons. 
With relations $C_{V,i} = 3k_B/(2 m_i)$, $C_{V,e} = 3k_B/(2m_e)$, and $ C_{V,i}\rho_i = C_{V,e} \rho_e$, we can get
\begin{equation}
\begin{aligned}
  C_{V,i}\rho_i \frac{\mathrm d \Delta T}{\mathrm d t}
=&
  W_\alpha \frac{T_e-32 \text{keV}}{T_e+32 \text{keV}} - 2W_{ie}\\
&- W_{m,i} + W_{m,e} + W_r + W_e
\label{eq:delta temperature}
\end{aligned}
\end{equation}
where $\Delta T = T_i-T_e$. In the isobaric model, $u_h \sim 0$, meaning that the expansion of ions and electrons does not contribute to energy loss. Consequently, both $W_{m,i}$ and $W_{m,e}$ are zero.
At the beginning of the ignition process, when the electron and ion temperature is sufficiently low, $W_\alpha \propto \langle \sigma v \rangle$, which highly depends on the ion temperature \cite{atzeni_inertial_2013}, $W_r \propto T_e ^{1/2}$ and $W_e \propto T_e^{7/2}$. Instead, $W_{ie} \propto \Delta T \cdot T_e^{-3/2}$, which means that at low temperatures, energy exchange between the ions and electrons dominates, as demonstrated in Fig.\ \ref{fig:stages}.
Also, due to the $\Delta T$ dependence and the large coefficient $A_{\omega,ei}$ in the ion-electron energy exchange power $W_{ie}$, it will dominate for a while before the temperature reaches a high level.
Consequently, the temperature difference between ion and electron starts to shrink, reaching a dynamic equilibrium. 
However, when the initial temperature difference is minimal (i.e., $f$ is close to 1), the process becomes less distinct, as it can be interpreted as having reached a state of dynamic equilibrium.

% {\color{red}Under verification}

% Having ignored all terms that do not dominate, we can derive the relaxation time $\tau_{ie}$ of the heating conduction between conduction. Note that, during the process, the $\alpha$ particle deposit energy and energy loss are low, so the electron and ion temperature can be seen as a constant.
%\begin{align}
%  \tau_{ie}
%=
%  \frac{C_{V,i} \rho_i T_e^{3/2}}{2 A_{\omega,ei} \rho_h^2 \ln \Lambda}
%\end{align}
%We also calculate the different powers in the relaxation process. We can find that the heating conduction is at least one order higher than other powers.

\subsection{Co-Heating as an Equilibrium Model}

As the ion and electron temperatures converge, they can be heated simultaneously under the influence of an increasing $\alpha$-particle heating power $W_\alpha$, as shown in B stage of Fig.\ \ref{fig:stages}. During this process, the ion-electron collisions maintain their close temperatures dynamically, by balancing the terms, e.g., $W_\alpha(f_{\alpha i} - f_{\alpha e})$, $2W_{ie}$, $W_r$ and $W_e$, in Eq.\ \eqref{eq:delta temperature}.
During the B stage depicted in 
Fig.\ \ref{fig:stages}, the $\alpha$-heating power significantly surpasses other energy loss powers, resulting in a co-heating process. However, as the electron temperature $T_e$ rises, the electron conduction power experiences a remarkable increase due to its $T_e^{7/2}$ dependency. When the electron temperature approaches approximately $\sim32\ \text{eV}$, $W_\alpha(f_{\alpha i} - f_{\alpha e})$ undergoes a sign change, indicating that the difference between the $\alpha$-particle heating contributions to the ion and electron temperatures is poised to reverse. These observations suggest that the balance maintained by the ion-electron collisions in Eq.\ \eqref{eq:delta temperature} may be disrupted.

\subsection{Bifurcated Heating}

In Stage C depicted in Fig.\ \ref{fig:stages}, the evolution of the ion and electron temperatures, $T_i$ and $T_e$, diverges significantly. This divergence can be attributed to the weakened ion-electron energy exchange power, $W_{ie}\propto \Delta T/(T_e)^{3/2}$ at high electron temperatures, as well as the differing energy loss and heating mechanisms between ions and electrons.

The heating effect on electron temperature is significantly suppressed, while ion temperature heating remains robust during Stage C.
For electrons, the significant enhancement in electron conduction loss, $W_e\propto T_e^{7/2}\sim 10^{28}$ strongly constrains the growth of the electron temperature, $T_e$. As evident in Eq.\ \eqref{eq:electron-temperature}, the $\alpha$-particle heating power, $W_\alpha\sim 10^{28}$ is effectively suppressed by $W_e$. Furthermore, the electron temperature $T_e$ increases at a rate proportional to $W_{ie}\propto \Delta T/(T_e)^{3/2}$, which is insufficient to keep up with the increasing temperatures. This trend is clearly illustrated in Fig.\ \ref{fig:stages}. Consequently, the electron temperature exhibits a slow growth ($\Delta T_e\sim 5$ keV) that can be disregarded, resulting in a gradual increase of $W_e$ towards a maximum value. Meanwhile $f_{\alpha i}$ and $f_{\alpha e}$ remain virtually unchanged, 
and $W_{ie}$ exhibits a nearly linear increase with $\propto \Delta T$. 
For ions, we delve into Eq.\ \eqref{eq:ion-temperature} and discover that they lack a significant energy loss mechanism besides transferring energy to electrons through collisions. Given the slow but steady increase in the relatively low energy exchange power $W_{ie}$, the heating effect from $\alpha$-particles, represented by $W_{\alpha i}= W_\alpha f_{\alpha i}$ dominates the ion temperature growth, resulting in a significant increase. Consequently, $T_i$ rises extremely rapidly, eventually pushing the ion-electron energy exchange power $W_{ie}$ to a notably high level. Meanwhile, the decreasing value of $f_\alpha$ is accompanied by a weakening of $W_\alpha$. 
To conclude, we refer to Eq.\ \eqref{eq:delta temperature}, which reveals that the difference in $\alpha$-particle heating,  $W_{\alpha} (f_{\alpha i}-f_{\alpha e}) \sim 10^{27}$, is significant. Additionally, the ion-electron energy exchange power, which is slowly increasing, contributes $2W_{ie}\sim 10^{26-27}$ to the overall heating process. However, the dominant factor is the electron conduction loss, estimated to be $W_{e} \sim 10^{28}$. Consequently, the ion and electron temperatures diverge significantly as they evolve.

\subsection{Reaching Saturated Temperatures}

During the stage D depicted in Fig.\ \ref{fig:stages}, a saturation phenomenon becomes evident in both ion and electron temperatures.
We have mentioned the slowly growing electron temperature $T_e$, electron conduction loss $W_e$, the increasing ion temperature $T_i$, ion-electron energy exchange $W_{ie}$ and the decreasing $\alpha$-particle heating $W_\alpha$ in stage C. 
With an increasing temperature gradient, the energy exchange $W_{ie}$ between ions and electrons significantly intensifies. This enhancement allows the energy exchange to be comparable with both the electron thermal conduction loss $W_e$ and the $\alpha$-particle heating $W_\alpha f_{\alpha e}$, ultimately, a balance is achieved between the energy loss and gain for electrons. Subsequently, the electron temperature nearly attains a state of saturation.
Later, with the increase of $T_i$, the $\alpha$-particle heating power decreases, while the ion-electron energy exchange $W_{ie}$ continues to adjust, maintaining the saturation of the electron temperature $T_e$. Eventually, the heating power from $\alpha$-particles, represented by $W_{\alpha}, W_{\alpha i}$ will decline to a certain low level, reaching a balance with the ion-electron conduction $W_{ie}$ as expressed in Eq.\ \eqref{eq:ion-temperature}. 
This balance contributes to the saturation of the ion temperature $T_i$. 

The intricate internal plasma environment poses significant non-linear constraints, stemming from the diverse energy gain and loss mechanisms at play. However, the ion-electron energy exchange, functioning as a self-regulating quantity, dynamically adjusts in response to the varying energy gain and loss mechanisms of ions and electrons, ultimately leading to the saturation phenomenon. This phenomenon serves as a vivid illustration of the complex non-linear constraints inherent in fusion plasmas.
Furthermore, our results underscore a non-isothermal phenomenon that plays a pivotal role in enhancing the efficient burning of hot-spots. Notably, the primary energy loss mechanism in hot-spots is closely linked to the electron temperature, whereas the fusion heating mechanism exhibits a positive correlation with ion temperature. This dynamic imbalance gives rise to a non-equilibrium hot-spot model, which emerges as an inevitable outcome and represents one of our most valuable predictions.

\section{Conclusion and discussions}
\label{sec:Conclusion and discussions}

In our research, we delve into a non-equilibrium model, extending it to consider both isobaric and isochoric conditions. These conditions exhibit variations in the densities, temperatures and expansion velocities of the hot spot. Our results reveal intriguing self-organization phenomena in ion and electron temperatures during the ignition process. Specifically, we find that ion temperature dominates over the electron temperature in this process. This phenomenon arises due to the significant heating effect caused by alpha particles, as well as the distinct deposition rates of alpha particle heating at high temperatures. Additionally, the reduced rate of energy exchange between electrons and D-T ions contributes to the observed bifurcation. During ignition, the inherent structure of higher ion temperature and lower electron temperature directly promotes the enhancement of the D-T reaction and reduces energy loss through electron conduction. Consequently, our ion-electron non-equilibrium model holds promise for improving inertial fusion ignition performed at current mega-joule laser facilities.

\begin{acknowledgments}
This work is supported by the Strategic Priority Research Program of Chinese Academy of Sciences (Grant Nos. XDA25010100 and XDA250050500), National Natural Science Foundation of China (Grants No. 12075204), and Shanghai Municipal Science and Technology Key Project (No. 22JC1401500). Dong Wu thanks the sponsorship from Yangyang Development Fund.

The junior undergraduate students, X.-Y. Fu, Z.-Y. Guo, Q.-H. Wang, and R.-C. Wang, all contributed equally to this work.
\end{acknowledgments}

% \begin{thebibliography}{99}
\bibliography{references}

\begin{thebibliography}{28}%
\makeatletter
\providecommand \@ifxundefined [1]{%
 \@ifx{#1\undefined}
}%
\providecommand \@ifnum [1]{%
 \ifnum #1\expandafter \@firstoftwo
 \else \expandafter \@secondoftwo
 \fi
}%
\providecommand \@ifx [1]{%
 \ifx #1\expandafter \@firstoftwo
 \else \expandafter \@secondoftwo
 \fi
}%
\providecommand \natexlab [1]{#1}%
\providecommand \enquote  [1]{``#1''}%
\providecommand \bibnamefont  [1]{#1}%
\providecommand \bibfnamefont [1]{#1}%
\providecommand \citenamefont [1]{#1}%
\providecommand \href@noop [0]{\@secondoftwo}%
\providecommand \href [0]{\begingroup \@sanitize@url \@href}%
\providecommand \@href[1]{\@@startlink{#1}\@@href}%
\providecommand \@@href[1]{\endgroup#1\@@endlink}%
\providecommand \@sanitize@url [0]{\catcode `\\12\catcode `\$12\catcode
  `\&12\catcode `\#12\catcode `\^12\catcode `\_12\catcode `\%12\relax}%
\providecommand \@@startlink[1]{}%
\providecommand \@@endlink[0]{}%
\providecommand \url  [0]{\begingroup\@sanitize@url \@url }%
\providecommand \@url [1]{\endgroup\@href {#1}{\urlprefix }}%
\providecommand \urlprefix  [0]{URL }%
\providecommand \Eprint [0]{\href }%
\providecommand \doibase [0]{http://dx.doi.org/}%
\providecommand \selectlanguage [0]{\@gobble}%
\providecommand \bibinfo  [0]{\@secondoftwo}%
\providecommand \bibfield  [0]{\@secondoftwo}%
\providecommand \translation [1]{[#1]}%
\providecommand \BibitemOpen [0]{}%
\providecommand \bibitemStop [0]{}%
\providecommand \bibitemNoStop [0]{.\EOS\space}%
\providecommand \EOS [0]{\spacefactor3000\relax}%
\providecommand \BibitemShut  [1]{\csname bibitem#1\endcsname}%
\let\auto@bib@innerbib\@empty
%</preamble>
\bibitem [{\citenamefont {Atzeni}()}]{atzeni_inertial_2013}%
  \BibitemOpen
  \bibfield  {author} {\bibinfo {author} {\bibfnamefont {S.}~\bibnamefont
  {Atzeni}},\ }in\ \href {\doibase 10.1007/978-3-319-00038-1_10} {\emph
  {\bibinfo {booktitle} {Laser-Plasma Interactions and Applications}}},\
  \bibinfo {editor} {edited by\ \bibinfo {editor} {\bibfnamefont
  {P.}~\bibnamefont {McKenna}}, \bibinfo {editor} {\bibfnamefont
  {D.}~\bibnamefont {Neely}}, \bibinfo {editor} {\bibfnamefont
  {R.}~\bibnamefont {Bingham}}, \ and\ \bibinfo {editor} {\bibfnamefont
  {D.}~\bibnamefont {Jaroszynski}}}\ (\bibinfo  {publisher} {Springer
  International Publishing})\ pp.\ \bibinfo {pages} {243--277}\BibitemShut
  {NoStop}%
\bibitem [{\citenamefont {Clavin}(2017)}]{clavin_quasi-isobaric_2017}%
  \BibitemOpen
  \bibfield  {author} {\bibinfo {author} {\bibfnamefont {P.}~\bibnamefont
  {Clavin}},\ }\href {\doibase 10.1016/j.combustflame.2016.05.019} {\bibfield
  {journal} {\bibinfo  {journal} {Combustion and Flame}\ }\bibinfo {series}
  {Special {Issue} in {Honor} of {Norbert} {Peters}},\ \textbf {\bibinfo
  {volume} {175}},\ \bibinfo {pages} {80} (\bibinfo {year} {2017})}\BibitemShut
  {NoStop}%
\bibitem [{\citenamefont {Rygg}\ \emph {et~al.}(2009)\citenamefont {Rygg},
  \citenamefont {Frenje}, \citenamefont {Li}, \citenamefont {Séguin},
  \citenamefont {Petrasso}, \citenamefont {Meyerhofer},\ and\ \citenamefont
  {Stoeckl}}]{rygg_electron-ion_2009}%
  \BibitemOpen
  \bibfield  {author} {\bibinfo {author} {\bibfnamefont {J.~R.}\ \bibnamefont
  {Rygg}}, \bibinfo {author} {\bibfnamefont {J.~A.}\ \bibnamefont {Frenje}},
  \bibinfo {author} {\bibfnamefont {C.~K.}\ \bibnamefont {Li}}, \bibinfo
  {author} {\bibfnamefont {F.~H.}\ \bibnamefont {Séguin}}, \bibinfo {author}
  {\bibfnamefont {R.~D.}\ \bibnamefont {Petrasso}}, \bibinfo {author}
  {\bibfnamefont {D.~D.}\ \bibnamefont {Meyerhofer}}, \ and\ \bibinfo {author}
  {\bibfnamefont {C.}~\bibnamefont {Stoeckl}},\ }\href {\doibase
  10.1103/PhysRevE.80.026403} {\bibfield  {journal} {\bibinfo  {journal}
  {Physical Review E}\ }\textbf {\bibinfo {volume} {80}},\ \bibinfo {pages}
  {026403} (\bibinfo {year} {2009})}\BibitemShut {NoStop}%
\bibitem [{\citenamefont {Tabak}\ \emph {et~al.}(2006)\citenamefont {Tabak},
  \citenamefont {Hinkel}, \citenamefont {Atzeni}, \citenamefont {Campbell},\
  and\ \citenamefont {Tanaka}}]{tabak_fast_2006}%
  \BibitemOpen
  \bibfield  {author} {\bibinfo {author} {\bibfnamefont {M.}~\bibnamefont
  {Tabak}}, \bibinfo {author} {\bibfnamefont {D.}~\bibnamefont {Hinkel}},
  \bibinfo {author} {\bibfnamefont {S.}~\bibnamefont {Atzeni}}, \bibinfo
  {author} {\bibfnamefont {E.~M.}\ \bibnamefont {Campbell}}, \ and\ \bibinfo
  {author} {\bibfnamefont {K.}~\bibnamefont {Tanaka}},\ }\href {\doibase
  10.13182/FST49-3-254} {\bibfield  {journal} {\bibinfo  {journal} {Fusion
  Science and Technology}\ }\textbf {\bibinfo {volume} {49}},\ \bibinfo {pages}
  {254} (\bibinfo {year} {2006})},\ \bibinfo {note} {publisher: Taylor \&
  Francis \_eprint: https://doi.org/10.13182/FST49-3-254}\BibitemShut {NoStop}%
\bibitem [{\citenamefont {Ghasemi}\ \emph {et~al.}(2014)\citenamefont
  {Ghasemi}, \citenamefont {Farahbod},\ and\ \citenamefont
  {Sobhanian}}]{ghasemi_analytical_2014}%
  \BibitemOpen
  \bibfield  {author} {\bibinfo {author} {\bibfnamefont {S.~A.}\ \bibnamefont
  {Ghasemi}}, \bibinfo {author} {\bibfnamefont {A.~H.}\ \bibnamefont
  {Farahbod}}, \ and\ \bibinfo {author} {\bibfnamefont {S.}~\bibnamefont
  {Sobhanian}},\ }\href {\doibase 10.1063/1.4891648} {\bibfield  {journal}
  {\bibinfo  {journal} {AIP Advances}\ }\textbf {\bibinfo {volume} {4}},\
  \bibinfo {pages} {077130} (\bibinfo {year} {2014})}\BibitemShut {NoStop}%
\bibitem [{\citenamefont {Xu}\ \emph {et~al.}(2023)\citenamefont {Xu},
  \citenamefont {Wu}, \citenamefont {Jiang}, \citenamefont {Kawata},\ and\
  \citenamefont {Zhang}}]{xu_formation_2023}%
  \BibitemOpen
  \bibfield  {author} {\bibinfo {author} {\bibfnamefont {Z.}~\bibnamefont
  {Xu}}, \bibinfo {author} {\bibfnamefont {F.}~\bibnamefont {Wu}}, \bibinfo
  {author} {\bibfnamefont {B.}~\bibnamefont {Jiang}}, \bibinfo {author}
  {\bibfnamefont {S.}~\bibnamefont {Kawata}}, \ and\ \bibinfo {author}
  {\bibfnamefont {J.}~\bibnamefont {Zhang}},\ }\href {\doibase
  10.1088/1741-4326/ad08e6} {\bibfield  {journal} {\bibinfo  {journal} {Nuclear
  Fusion}\ }\textbf {\bibinfo {volume} {63}},\ \bibinfo {pages} {126062}
  (\bibinfo {year} {2023})},\ \bibinfo {note} {publisher: IOP
  Publishing}\BibitemShut {NoStop}%
\bibitem [{\citenamefont {Clark}\ and\ \citenamefont
  {Tabak}(2007)}]{clark_self-similar_2007-1}%
  \BibitemOpen
  \bibfield  {author} {\bibinfo {author} {\bibfnamefont {D.~S.}\ \bibnamefont
  {Clark}}\ and\ \bibinfo {author} {\bibfnamefont {M.}~\bibnamefont {Tabak}},\
  }\href {\doibase 10.1088/0029-5515/47/9/011} {\bibfield  {journal} {\bibinfo
  {journal} {Nuclear Fusion}\ }\textbf {\bibinfo {volume} {47}},\ \bibinfo
  {pages} {1147} (\bibinfo {year} {2007})}\BibitemShut {NoStop}%
\bibitem [{\citenamefont {Farahbod}\ \emph {et~al.}(2014)\citenamefont
  {Farahbod}, \citenamefont {Ghasemi}, \citenamefont {Jafari}, \citenamefont
  {Rezaei},\ and\ \citenamefont {Sobhanian}}]{farahbod_improvement_2014}%
  \BibitemOpen
  \bibfield  {author} {\bibinfo {author} {\bibfnamefont {A.~H.}\ \bibnamefont
  {Farahbod}}, \bibinfo {author} {\bibfnamefont {S.~A.}\ \bibnamefont
  {Ghasemi}}, \bibinfo {author} {\bibfnamefont {M.~J.}\ \bibnamefont {Jafari}},
  \bibinfo {author} {\bibfnamefont {S.}~\bibnamefont {Rezaei}}, \ and\ \bibinfo
  {author} {\bibfnamefont {S.}~\bibnamefont {Sobhanian}},\ }\href {\doibase
  10.1140/epjd/e2014-50353-6} {\bibfield  {journal} {\bibinfo  {journal} {The
  European Physical Journal D}\ }\textbf {\bibinfo {volume} {68}},\ \bibinfo
  {pages} {314} (\bibinfo {year} {2014})}\BibitemShut {NoStop}%
\bibitem [{\citenamefont {Zylstra}\ \emph {et~al.}(2022)\citenamefont
  {Zylstra}, \citenamefont {Hurricane}, \citenamefont {Callahan}, \citenamefont
  {Kritcher}, \citenamefont {Ralph}, \citenamefont {Robey}, \citenamefont
  {Ross}, \citenamefont {Young}, \citenamefont {Baker}, \citenamefont {Casey},
  \citenamefont {Döppner}, \citenamefont {Divol}, \citenamefont {Hohenberger},
  \citenamefont {Le~Pape}, \citenamefont {Pak}, \citenamefont {Patel},
  \citenamefont {Tommasini}, \citenamefont {Ali}, \citenamefont {Amendt},
  \citenamefont {Atherton}, \citenamefont {Bachmann}, \citenamefont {Bailey},
  \citenamefont {Benedetti}, \citenamefont {Berzak~Hopkins}, \citenamefont
  {Betti}, \citenamefont {Bhandarkar}, \citenamefont {Biener}, \citenamefont
  {Bionta}, \citenamefont {Birge}, \citenamefont {Bond}, \citenamefont
  {Bradley}, \citenamefont {Braun}, \citenamefont {Briggs}, \citenamefont
  {Bruhn}, \citenamefont {Celliers}, \citenamefont {Chang}, \citenamefont
  {Chapman}, \citenamefont {Chen}, \citenamefont {Choate}, \citenamefont
  {Christopherson}, \citenamefont {Clark}, \citenamefont {Crippen},
  \citenamefont {Dewald}, \citenamefont {Dittrich}, \citenamefont {Edwards},
  \citenamefont {Farmer}, \citenamefont {Field}, \citenamefont {Fittinghoff},
  \citenamefont {Frenje}, \citenamefont {Gaffney}, \citenamefont
  {Gatu~Johnson}, \citenamefont {Glenzer}, \citenamefont {Grim}, \citenamefont
  {Haan}, \citenamefont {Hahn}, \citenamefont {Hall}, \citenamefont {Hammel},
  \citenamefont {Harte}, \citenamefont {Hartouni}, \citenamefont {Heebner},
  \citenamefont {Hernandez}, \citenamefont {Herrmann}, \citenamefont
  {Herrmann}, \citenamefont {Hinkel}, \citenamefont {Ho}, \citenamefont
  {Holder}, \citenamefont {Hsing}, \citenamefont {Huang}, \citenamefont
  {Humbird}, \citenamefont {Izumi}, \citenamefont {Jarrott}, \citenamefont
  {Jeet}, \citenamefont {Jones}, \citenamefont {Kerbel}, \citenamefont {Kerr},
  \citenamefont {Khan}, \citenamefont {Kilkenny}, \citenamefont {Kim},
  \citenamefont {Geppert~Kleinrath}, \citenamefont {Geppert~Kleinrath},
  \citenamefont {Kong}, \citenamefont {Koning}, \citenamefont {Kroll},
  \citenamefont {Kruse}, \citenamefont {Kustowski}, \citenamefont {Landen},
  \citenamefont {Langer}, \citenamefont {Larson}, \citenamefont {Lemos},
  \citenamefont {Lindl}, \citenamefont {Ma}, \citenamefont {MacDonald},
  \citenamefont {MacGowan}, \citenamefont {Mackinnon}, \citenamefont
  {MacLaren}, \citenamefont {MacPhee}, \citenamefont {Marinak}, \citenamefont
  {Mariscal}, \citenamefont {Marley}, \citenamefont {Masse}, \citenamefont
  {Meaney}, \citenamefont {Meezan}, \citenamefont {Michel}, \citenamefont
  {Millot}, \citenamefont {Milovich}, \citenamefont {Moody}, \citenamefont
  {Moore}, \citenamefont {Morton}, \citenamefont {Murphy}, \citenamefont
  {Newman}, \citenamefont {Di~Nicola}, \citenamefont {Nikroo}, \citenamefont
  {Nora}, \citenamefont {Patel}, \citenamefont {Pelz}, \citenamefont
  {Peterson}, \citenamefont {Ping}, \citenamefont {Pollock}, \citenamefont
  {Ratledge}, \citenamefont {Rice}, \citenamefont {Rinderknecht}, \citenamefont
  {Rosen}, \citenamefont {Rubery}, \citenamefont {Salmonson}, \citenamefont
  {Sater}, \citenamefont {Schiaffino}, \citenamefont {Schlossberg},
  \citenamefont {Schneider}, \citenamefont {Schroeder}, \citenamefont {Scott},
  \citenamefont {Sepke}, \citenamefont {Sequoia}, \citenamefont {Sherlock},
  \citenamefont {Shin}, \citenamefont {Smalyuk}, \citenamefont {Spears},
  \citenamefont {Springer}, \citenamefont {Stadermann}, \citenamefont
  {Stoupin}, \citenamefont {Strozzi}, \citenamefont {Suter}, \citenamefont
  {Thomas}, \citenamefont {Town}, \citenamefont {Tubman}, \citenamefont
  {Trosseille}, \citenamefont {Volegov}, \citenamefont {Weber}, \citenamefont
  {Widmann}, \citenamefont {Wild}, \citenamefont {Wilde}, \citenamefont
  {Van~Wonterghem}, \citenamefont {Woods}, \citenamefont {Woodworth},
  \citenamefont {Yamaguchi}, \citenamefont {Yang},\ and\ \citenamefont
  {Zimmerman}}]{zylstra_burning_2022}%
  \BibitemOpen
  \bibfield  {author} {\bibinfo {author} {\bibfnamefont {A.~B.}\ \bibnamefont
  {Zylstra}}, \bibinfo {author} {\bibfnamefont {O.~A.}\ \bibnamefont
  {Hurricane}}, \bibinfo {author} {\bibfnamefont {D.~A.}\ \bibnamefont
  {Callahan}}, \bibinfo {author} {\bibfnamefont {A.~L.}\ \bibnamefont
  {Kritcher}}, \bibinfo {author} {\bibfnamefont {J.~E.}\ \bibnamefont {Ralph}},
  \bibinfo {author} {\bibfnamefont {H.~F.}\ \bibnamefont {Robey}}, \bibinfo
  {author} {\bibfnamefont {J.~S.}\ \bibnamefont {Ross}}, \bibinfo {author}
  {\bibfnamefont {C.~V.}\ \bibnamefont {Young}}, \bibinfo {author}
  {\bibfnamefont {K.~L.}\ \bibnamefont {Baker}}, \bibinfo {author}
  {\bibfnamefont {D.~T.}\ \bibnamefont {Casey}}, \bibinfo {author}
  {\bibfnamefont {T.}~\bibnamefont {Döppner}}, \bibinfo {author}
  {\bibfnamefont {L.}~\bibnamefont {Divol}}, \bibinfo {author} {\bibfnamefont
  {M.}~\bibnamefont {Hohenberger}}, \bibinfo {author} {\bibfnamefont
  {S.}~\bibnamefont {Le~Pape}}, \bibinfo {author} {\bibfnamefont
  {A.}~\bibnamefont {Pak}}, \bibinfo {author} {\bibfnamefont {P.~K.}\
  \bibnamefont {Patel}}, \bibinfo {author} {\bibfnamefont {R.}~\bibnamefont
  {Tommasini}}, \bibinfo {author} {\bibfnamefont {S.~J.}\ \bibnamefont {Ali}},
  \bibinfo {author} {\bibfnamefont {P.~A.}\ \bibnamefont {Amendt}}, \bibinfo
  {author} {\bibfnamefont {L.~J.}\ \bibnamefont {Atherton}}, \bibinfo {author}
  {\bibfnamefont {B.}~\bibnamefont {Bachmann}}, \bibinfo {author}
  {\bibfnamefont {D.}~\bibnamefont {Bailey}}, \bibinfo {author} {\bibfnamefont
  {L.~R.}\ \bibnamefont {Benedetti}}, \bibinfo {author} {\bibfnamefont
  {L.}~\bibnamefont {Berzak~Hopkins}}, \bibinfo {author} {\bibfnamefont
  {R.}~\bibnamefont {Betti}}, \bibinfo {author} {\bibfnamefont {S.~D.}\
  \bibnamefont {Bhandarkar}}, \bibinfo {author} {\bibfnamefont
  {J.}~\bibnamefont {Biener}}, \bibinfo {author} {\bibfnamefont {R.~M.}\
  \bibnamefont {Bionta}}, \bibinfo {author} {\bibfnamefont {N.~W.}\
  \bibnamefont {Birge}}, \bibinfo {author} {\bibfnamefont {E.~J.}\ \bibnamefont
  {Bond}}, \bibinfo {author} {\bibfnamefont {D.~K.}\ \bibnamefont {Bradley}},
  \bibinfo {author} {\bibfnamefont {T.}~\bibnamefont {Braun}}, \bibinfo
  {author} {\bibfnamefont {T.~M.}\ \bibnamefont {Briggs}}, \bibinfo {author}
  {\bibfnamefont {M.~W.}\ \bibnamefont {Bruhn}}, \bibinfo {author}
  {\bibfnamefont {P.~M.}\ \bibnamefont {Celliers}}, \bibinfo {author}
  {\bibfnamefont {B.}~\bibnamefont {Chang}}, \bibinfo {author} {\bibfnamefont
  {T.}~\bibnamefont {Chapman}}, \bibinfo {author} {\bibfnamefont
  {H.}~\bibnamefont {Chen}}, \bibinfo {author} {\bibfnamefont {C.}~\bibnamefont
  {Choate}}, \bibinfo {author} {\bibfnamefont {A.~R.}\ \bibnamefont
  {Christopherson}}, \bibinfo {author} {\bibfnamefont {D.~S.}\ \bibnamefont
  {Clark}}, \bibinfo {author} {\bibfnamefont {J.~W.}\ \bibnamefont {Crippen}},
  \bibinfo {author} {\bibfnamefont {E.~L.}\ \bibnamefont {Dewald}}, \bibinfo
  {author} {\bibfnamefont {T.~R.}\ \bibnamefont {Dittrich}}, \bibinfo {author}
  {\bibfnamefont {M.~J.}\ \bibnamefont {Edwards}}, \bibinfo {author}
  {\bibfnamefont {W.~A.}\ \bibnamefont {Farmer}}, \bibinfo {author}
  {\bibfnamefont {J.~E.}\ \bibnamefont {Field}}, \bibinfo {author}
  {\bibfnamefont {D.}~\bibnamefont {Fittinghoff}}, \bibinfo {author}
  {\bibfnamefont {J.}~\bibnamefont {Frenje}}, \bibinfo {author} {\bibfnamefont
  {J.}~\bibnamefont {Gaffney}}, \bibinfo {author} {\bibfnamefont
  {M.}~\bibnamefont {Gatu~Johnson}}, \bibinfo {author} {\bibfnamefont {S.~H.}\
  \bibnamefont {Glenzer}}, \bibinfo {author} {\bibfnamefont {G.~P.}\
  \bibnamefont {Grim}}, \bibinfo {author} {\bibfnamefont {S.}~\bibnamefont
  {Haan}}, \bibinfo {author} {\bibfnamefont {K.~D.}\ \bibnamefont {Hahn}},
  \bibinfo {author} {\bibfnamefont {G.~N.}\ \bibnamefont {Hall}}, \bibinfo
  {author} {\bibfnamefont {B.~A.}\ \bibnamefont {Hammel}}, \bibinfo {author}
  {\bibfnamefont {J.}~\bibnamefont {Harte}}, \bibinfo {author} {\bibfnamefont
  {E.}~\bibnamefont {Hartouni}}, \bibinfo {author} {\bibfnamefont {J.~E.}\
  \bibnamefont {Heebner}}, \bibinfo {author} {\bibfnamefont {V.~J.}\
  \bibnamefont {Hernandez}}, \bibinfo {author} {\bibfnamefont {H.}~\bibnamefont
  {Herrmann}}, \bibinfo {author} {\bibfnamefont {M.~C.}\ \bibnamefont
  {Herrmann}}, \bibinfo {author} {\bibfnamefont {D.~E.}\ \bibnamefont
  {Hinkel}}, \bibinfo {author} {\bibfnamefont {D.~D.}\ \bibnamefont {Ho}},
  \bibinfo {author} {\bibfnamefont {J.~P.}\ \bibnamefont {Holder}}, \bibinfo
  {author} {\bibfnamefont {W.~W.}\ \bibnamefont {Hsing}}, \bibinfo {author}
  {\bibfnamefont {H.}~\bibnamefont {Huang}}, \bibinfo {author} {\bibfnamefont
  {K.~D.}\ \bibnamefont {Humbird}}, \bibinfo {author} {\bibfnamefont
  {N.}~\bibnamefont {Izumi}}, \bibinfo {author} {\bibfnamefont {L.~C.}\
  \bibnamefont {Jarrott}}, \bibinfo {author} {\bibfnamefont {J.}~\bibnamefont
  {Jeet}}, \bibinfo {author} {\bibfnamefont {O.}~\bibnamefont {Jones}},
  \bibinfo {author} {\bibfnamefont {G.~D.}\ \bibnamefont {Kerbel}}, \bibinfo
  {author} {\bibfnamefont {S.~M.}\ \bibnamefont {Kerr}}, \bibinfo {author}
  {\bibfnamefont {S.~F.}\ \bibnamefont {Khan}}, \bibinfo {author}
  {\bibfnamefont {J.}~\bibnamefont {Kilkenny}}, \bibinfo {author}
  {\bibfnamefont {Y.}~\bibnamefont {Kim}}, \bibinfo {author} {\bibfnamefont
  {H.}~\bibnamefont {Geppert~Kleinrath}}, \bibinfo {author} {\bibfnamefont
  {V.}~\bibnamefont {Geppert~Kleinrath}}, \bibinfo {author} {\bibfnamefont
  {C.}~\bibnamefont {Kong}}, \bibinfo {author} {\bibfnamefont {J.~M.}\
  \bibnamefont {Koning}}, \bibinfo {author} {\bibfnamefont {J.~J.}\
  \bibnamefont {Kroll}}, \bibinfo {author} {\bibfnamefont {M.~K.~G.}\
  \bibnamefont {Kruse}}, \bibinfo {author} {\bibfnamefont {B.}~\bibnamefont
  {Kustowski}}, \bibinfo {author} {\bibfnamefont {O.~L.}\ \bibnamefont
  {Landen}}, \bibinfo {author} {\bibfnamefont {S.}~\bibnamefont {Langer}},
  \bibinfo {author} {\bibfnamefont {D.}~\bibnamefont {Larson}}, \bibinfo
  {author} {\bibfnamefont {N.~C.}\ \bibnamefont {Lemos}}, \bibinfo {author}
  {\bibfnamefont {J.~D.}\ \bibnamefont {Lindl}}, \bibinfo {author}
  {\bibfnamefont {T.}~\bibnamefont {Ma}}, \bibinfo {author} {\bibfnamefont
  {M.~J.}\ \bibnamefont {MacDonald}}, \bibinfo {author} {\bibfnamefont {B.~J.}\
  \bibnamefont {MacGowan}}, \bibinfo {author} {\bibfnamefont {A.~J.}\
  \bibnamefont {Mackinnon}}, \bibinfo {author} {\bibfnamefont {S.~A.}\
  \bibnamefont {MacLaren}}, \bibinfo {author} {\bibfnamefont {A.~G.}\
  \bibnamefont {MacPhee}}, \bibinfo {author} {\bibfnamefont {M.~M.}\
  \bibnamefont {Marinak}}, \bibinfo {author} {\bibfnamefont {D.~A.}\
  \bibnamefont {Mariscal}}, \bibinfo {author} {\bibfnamefont {E.~V.}\
  \bibnamefont {Marley}}, \bibinfo {author} {\bibfnamefont {L.}~\bibnamefont
  {Masse}}, \bibinfo {author} {\bibfnamefont {K.}~\bibnamefont {Meaney}},
  \bibinfo {author} {\bibfnamefont {N.~B.}\ \bibnamefont {Meezan}}, \bibinfo
  {author} {\bibfnamefont {P.~A.}\ \bibnamefont {Michel}}, \bibinfo {author}
  {\bibfnamefont {M.}~\bibnamefont {Millot}}, \bibinfo {author} {\bibfnamefont
  {J.~L.}\ \bibnamefont {Milovich}}, \bibinfo {author} {\bibfnamefont {J.~D.}\
  \bibnamefont {Moody}}, \bibinfo {author} {\bibfnamefont {A.~S.}\ \bibnamefont
  {Moore}}, \bibinfo {author} {\bibfnamefont {J.~W.}\ \bibnamefont {Morton}},
  \bibinfo {author} {\bibfnamefont {T.}~\bibnamefont {Murphy}}, \bibinfo
  {author} {\bibfnamefont {K.}~\bibnamefont {Newman}}, \bibinfo {author}
  {\bibfnamefont {J.-M.~G.}\ \bibnamefont {Di~Nicola}}, \bibinfo {author}
  {\bibfnamefont {A.}~\bibnamefont {Nikroo}}, \bibinfo {author} {\bibfnamefont
  {R.}~\bibnamefont {Nora}}, \bibinfo {author} {\bibfnamefont {M.~V.}\
  \bibnamefont {Patel}}, \bibinfo {author} {\bibfnamefont {L.~J.}\ \bibnamefont
  {Pelz}}, \bibinfo {author} {\bibfnamefont {J.~L.}\ \bibnamefont {Peterson}},
  \bibinfo {author} {\bibfnamefont {Y.}~\bibnamefont {Ping}}, \bibinfo {author}
  {\bibfnamefont {B.~B.}\ \bibnamefont {Pollock}}, \bibinfo {author}
  {\bibfnamefont {M.}~\bibnamefont {Ratledge}}, \bibinfo {author}
  {\bibfnamefont {N.~G.}\ \bibnamefont {Rice}}, \bibinfo {author}
  {\bibfnamefont {H.}~\bibnamefont {Rinderknecht}}, \bibinfo {author}
  {\bibfnamefont {M.}~\bibnamefont {Rosen}}, \bibinfo {author} {\bibfnamefont
  {M.~S.}\ \bibnamefont {Rubery}}, \bibinfo {author} {\bibfnamefont {J.~D.}\
  \bibnamefont {Salmonson}}, \bibinfo {author} {\bibfnamefont {J.}~\bibnamefont
  {Sater}}, \bibinfo {author} {\bibfnamefont {S.}~\bibnamefont {Schiaffino}},
  \bibinfo {author} {\bibfnamefont {D.~J.}\ \bibnamefont {Schlossberg}},
  \bibinfo {author} {\bibfnamefont {M.~B.}\ \bibnamefont {Schneider}}, \bibinfo
  {author} {\bibfnamefont {C.~R.}\ \bibnamefont {Schroeder}}, \bibinfo {author}
  {\bibfnamefont {H.~A.}\ \bibnamefont {Scott}}, \bibinfo {author}
  {\bibfnamefont {S.~M.}\ \bibnamefont {Sepke}}, \bibinfo {author}
  {\bibfnamefont {K.}~\bibnamefont {Sequoia}}, \bibinfo {author} {\bibfnamefont
  {M.~W.}\ \bibnamefont {Sherlock}}, \bibinfo {author} {\bibfnamefont
  {S.}~\bibnamefont {Shin}}, \bibinfo {author} {\bibfnamefont {V.~A.}\
  \bibnamefont {Smalyuk}}, \bibinfo {author} {\bibfnamefont {B.~K.}\
  \bibnamefont {Spears}}, \bibinfo {author} {\bibfnamefont {P.~T.}\
  \bibnamefont {Springer}}, \bibinfo {author} {\bibfnamefont {M.}~\bibnamefont
  {Stadermann}}, \bibinfo {author} {\bibfnamefont {S.}~\bibnamefont {Stoupin}},
  \bibinfo {author} {\bibfnamefont {D.~J.}\ \bibnamefont {Strozzi}}, \bibinfo
  {author} {\bibfnamefont {L.~J.}\ \bibnamefont {Suter}}, \bibinfo {author}
  {\bibfnamefont {C.~A.}\ \bibnamefont {Thomas}}, \bibinfo {author}
  {\bibfnamefont {R.~P.~J.}\ \bibnamefont {Town}}, \bibinfo {author}
  {\bibfnamefont {E.~R.}\ \bibnamefont {Tubman}}, \bibinfo {author}
  {\bibfnamefont {C.}~\bibnamefont {Trosseille}}, \bibinfo {author}
  {\bibfnamefont {P.~L.}\ \bibnamefont {Volegov}}, \bibinfo {author}
  {\bibfnamefont {C.~R.}\ \bibnamefont {Weber}}, \bibinfo {author}
  {\bibfnamefont {K.}~\bibnamefont {Widmann}}, \bibinfo {author} {\bibfnamefont
  {C.}~\bibnamefont {Wild}}, \bibinfo {author} {\bibfnamefont {C.~H.}\
  \bibnamefont {Wilde}}, \bibinfo {author} {\bibfnamefont {B.~M.}\ \bibnamefont
  {Van~Wonterghem}}, \bibinfo {author} {\bibfnamefont {D.~T.}\ \bibnamefont
  {Woods}}, \bibinfo {author} {\bibfnamefont {B.~N.}\ \bibnamefont
  {Woodworth}}, \bibinfo {author} {\bibfnamefont {M.}~\bibnamefont
  {Yamaguchi}}, \bibinfo {author} {\bibfnamefont {S.~T.}\ \bibnamefont {Yang}},
  \ and\ \bibinfo {author} {\bibfnamefont {G.~B.}\ \bibnamefont {Zimmerman}},\
  }\href {\doibase 10.1038/s41586-021-04281-w} {\bibfield  {journal} {\bibinfo
  {journal} {Nature}\ }\textbf {\bibinfo {volume} {601}},\ \bibinfo {pages}
  {542} (\bibinfo {year} {2022})},\ \bibinfo {note} {number: 7894 Publisher:
  Nature Publishing Group}\BibitemShut {NoStop}%
\bibitem [{\citenamefont {Acree}\ \emph {et~al.}(2022)\citenamefont {Acree},
  \citenamefont {Abu-Shawareb},\ and\ \citenamefont
  {et~al}}]{acree_lawson_2022}%
  \BibitemOpen
  \bibfield  {author} {\bibinfo {author} {\bibfnamefont {R.}~\bibnamefont
  {Acree}}, \bibinfo {author} {\bibfnamefont {H.}~\bibnamefont {Abu-Shawareb}},
  \ and\ \bibinfo {author} {\bibnamefont {et~al}},\ }\href {\doibase
  10.1103/PhysRevLett.129.075001} {\bibfield  {journal} {\bibinfo  {journal}
  {Physical Review Letters}\ }\textbf {\bibinfo {volume} {129}},\ \bibinfo
  {pages} {075001} (\bibinfo {year} {2022})}\BibitemShut {NoStop}%
\bibitem [{\citenamefont {Acree}\ \emph {et~al.}(2024)\citenamefont {Acree},
  \citenamefont {Abu-Shawareb},\ and\ \citenamefont
  {et~al}}]{acree_achievement_2024}%
  \BibitemOpen
  \bibfield  {author} {\bibinfo {author} {\bibfnamefont {R.}~\bibnamefont
  {Acree}}, \bibinfo {author} {\bibfnamefont {H.}~\bibnamefont {Abu-Shawareb}},
  \ and\ \bibinfo {author} {\bibnamefont {et~al}},\ }\href {\doibase
  10.1103/PhysRevLett.132.065102} {\bibfield  {journal} {\bibinfo  {journal}
  {Physical Review Letters}\ }\textbf {\bibinfo {volume} {132}},\ \bibinfo
  {pages} {065102} (\bibinfo {year} {2024})}\BibitemShut {NoStop}%
\bibitem [{\citenamefont {Hurricane}\ \emph {et~al.}(2014)\citenamefont
  {Hurricane}, \citenamefont {Callahan}, \citenamefont {Casey}, \citenamefont
  {Celliers}, \citenamefont {Cerjan}, \citenamefont {Dewald}, \citenamefont
  {Dittrich}, \citenamefont {Döppner}, \citenamefont {Hinkel}, \citenamefont
  {Hopkins}, \citenamefont {Kline}, \citenamefont {Le~Pape}, \citenamefont
  {Ma}, \citenamefont {MacPhee}, \citenamefont {Milovich}, \citenamefont {Pak},
  \citenamefont {Park}, \citenamefont {Patel}, \citenamefont {Remington},
  \citenamefont {Salmonson}, \citenamefont {Springer},\ and\ \citenamefont
  {Tommasini}}]{hurricane_fuel_2014}%
  \BibitemOpen
  \bibfield  {author} {\bibinfo {author} {\bibfnamefont {O.~A.}\ \bibnamefont
  {Hurricane}}, \bibinfo {author} {\bibfnamefont {D.~A.}\ \bibnamefont
  {Callahan}}, \bibinfo {author} {\bibfnamefont {D.~T.}\ \bibnamefont {Casey}},
  \bibinfo {author} {\bibfnamefont {P.~M.}\ \bibnamefont {Celliers}}, \bibinfo
  {author} {\bibfnamefont {C.}~\bibnamefont {Cerjan}}, \bibinfo {author}
  {\bibfnamefont {E.~L.}\ \bibnamefont {Dewald}}, \bibinfo {author}
  {\bibfnamefont {T.~R.}\ \bibnamefont {Dittrich}}, \bibinfo {author}
  {\bibfnamefont {T.}~\bibnamefont {Döppner}}, \bibinfo {author}
  {\bibfnamefont {D.~E.}\ \bibnamefont {Hinkel}}, \bibinfo {author}
  {\bibfnamefont {L.~F.~B.}\ \bibnamefont {Hopkins}}, \bibinfo {author}
  {\bibfnamefont {J.~L.}\ \bibnamefont {Kline}}, \bibinfo {author}
  {\bibfnamefont {S.}~\bibnamefont {Le~Pape}}, \bibinfo {author} {\bibfnamefont
  {T.}~\bibnamefont {Ma}}, \bibinfo {author} {\bibfnamefont {A.~G.}\
  \bibnamefont {MacPhee}}, \bibinfo {author} {\bibfnamefont {J.~L.}\
  \bibnamefont {Milovich}}, \bibinfo {author} {\bibfnamefont {A.}~\bibnamefont
  {Pak}}, \bibinfo {author} {\bibfnamefont {H.-S.}\ \bibnamefont {Park}},
  \bibinfo {author} {\bibfnamefont {P.~K.}\ \bibnamefont {Patel}}, \bibinfo
  {author} {\bibfnamefont {B.~A.}\ \bibnamefont {Remington}}, \bibinfo {author}
  {\bibfnamefont {J.~D.}\ \bibnamefont {Salmonson}}, \bibinfo {author}
  {\bibfnamefont {P.~T.}\ \bibnamefont {Springer}}, \ and\ \bibinfo {author}
  {\bibfnamefont {R.}~\bibnamefont {Tommasini}},\ }\href {\doibase
  10.1038/nature13008} {\bibfield  {journal} {\bibinfo  {journal} {Nature}\
  }\textbf {\bibinfo {volume} {506}},\ \bibinfo {pages} {343} (\bibinfo {year}
  {2014})},\ \bibinfo {note} {publisher: Nature Publishing Group}\BibitemShut
  {NoStop}%
\bibitem [{\citenamefont {Lindl}\ \emph {et~al.}(2014)\citenamefont {Lindl},
  \citenamefont {Landen}, \citenamefont {Edwards}, \citenamefont {Moses},\ and\
  \citenamefont {{NIC Team}}}]{lindl_review_2014}%
  \BibitemOpen
  \bibfield  {author} {\bibinfo {author} {\bibfnamefont {J.}~\bibnamefont
  {Lindl}}, \bibinfo {author} {\bibfnamefont {O.}~\bibnamefont {Landen}},
  \bibinfo {author} {\bibfnamefont {J.}~\bibnamefont {Edwards}}, \bibinfo
  {author} {\bibfnamefont {E.}~\bibnamefont {Moses}}, \ and\ \bibinfo {author}
  {\bibnamefont {{NIC Team}}},\ }\href {\doibase 10.1063/1.4865400} {\bibfield
  {journal} {\bibinfo  {journal} {Physics of Plasmas}\ }\textbf {\bibinfo
  {volume} {21}},\ \bibinfo {pages} {020501} (\bibinfo {year}
  {2014})}\BibitemShut {NoStop}%
\bibitem [{\citenamefont {Hartouni}\ \emph {et~al.}(2023)\citenamefont
  {Hartouni}, \citenamefont {Moore}, \citenamefont {Crilly}, \citenamefont
  {Appelbe}, \citenamefont {Amendt}, \citenamefont {Baker}, \citenamefont
  {Casey}, \citenamefont {Clark}, \citenamefont {Döppner}, \citenamefont
  {Eckart}, \citenamefont {Field}, \citenamefont {Gatu-Johnson}, \citenamefont
  {Grim}, \citenamefont {Hatarik}, \citenamefont {Jeet}, \citenamefont {Kerr},
  \citenamefont {Kilkenny}, \citenamefont {Kritcher}, \citenamefont {Meaney},
  \citenamefont {Milovich}, \citenamefont {Munro}, \citenamefont {Nora},
  \citenamefont {Pak}, \citenamefont {Ralph}, \citenamefont {Robey},
  \citenamefont {Ross}, \citenamefont {Schlossberg}, \citenamefont {Sepke},
  \citenamefont {Spears}, \citenamefont {Young},\ and\ \citenamefont
  {Zylstra}}]{hartouni_evidence_2023}%
  \BibitemOpen
  \bibfield  {author} {\bibinfo {author} {\bibfnamefont {E.~P.}\ \bibnamefont
  {Hartouni}}, \bibinfo {author} {\bibfnamefont {A.~S.}\ \bibnamefont {Moore}},
  \bibinfo {author} {\bibfnamefont {A.~J.}\ \bibnamefont {Crilly}}, \bibinfo
  {author} {\bibfnamefont {B.~D.}\ \bibnamefont {Appelbe}}, \bibinfo {author}
  {\bibfnamefont {P.~A.}\ \bibnamefont {Amendt}}, \bibinfo {author}
  {\bibfnamefont {K.~L.}\ \bibnamefont {Baker}}, \bibinfo {author}
  {\bibfnamefont {D.~T.}\ \bibnamefont {Casey}}, \bibinfo {author}
  {\bibfnamefont {D.~S.}\ \bibnamefont {Clark}}, \bibinfo {author}
  {\bibfnamefont {T.}~\bibnamefont {Döppner}}, \bibinfo {author}
  {\bibfnamefont {M.~J.}\ \bibnamefont {Eckart}}, \bibinfo {author}
  {\bibfnamefont {J.~E.}\ \bibnamefont {Field}}, \bibinfo {author}
  {\bibfnamefont {M.}~\bibnamefont {Gatu-Johnson}}, \bibinfo {author}
  {\bibfnamefont {G.~P.}\ \bibnamefont {Grim}}, \bibinfo {author}
  {\bibfnamefont {R.}~\bibnamefont {Hatarik}}, \bibinfo {author} {\bibfnamefont
  {J.}~\bibnamefont {Jeet}}, \bibinfo {author} {\bibfnamefont {S.~M.}\
  \bibnamefont {Kerr}}, \bibinfo {author} {\bibfnamefont {J.}~\bibnamefont
  {Kilkenny}}, \bibinfo {author} {\bibfnamefont {A.~L.}\ \bibnamefont
  {Kritcher}}, \bibinfo {author} {\bibfnamefont {K.~D.}\ \bibnamefont
  {Meaney}}, \bibinfo {author} {\bibfnamefont {J.~L.}\ \bibnamefont
  {Milovich}}, \bibinfo {author} {\bibfnamefont {D.~H.}\ \bibnamefont {Munro}},
  \bibinfo {author} {\bibfnamefont {R.~C.}\ \bibnamefont {Nora}}, \bibinfo
  {author} {\bibfnamefont {A.~E.}\ \bibnamefont {Pak}}, \bibinfo {author}
  {\bibfnamefont {J.~E.}\ \bibnamefont {Ralph}}, \bibinfo {author}
  {\bibfnamefont {H.~F.}\ \bibnamefont {Robey}}, \bibinfo {author}
  {\bibfnamefont {J.~S.}\ \bibnamefont {Ross}}, \bibinfo {author}
  {\bibfnamefont {D.~J.}\ \bibnamefont {Schlossberg}}, \bibinfo {author}
  {\bibfnamefont {S.~M.}\ \bibnamefont {Sepke}}, \bibinfo {author}
  {\bibfnamefont {B.~K.}\ \bibnamefont {Spears}}, \bibinfo {author}
  {\bibfnamefont {C.~V.}\ \bibnamefont {Young}}, \ and\ \bibinfo {author}
  {\bibfnamefont {A.~B.}\ \bibnamefont {Zylstra}},\ }\href {\doibase
  10.1038/s41567-022-01809-3} {\bibfield  {journal} {\bibinfo  {journal}
  {Nature Physics}\ }\textbf {\bibinfo {volume} {19}},\ \bibinfo {pages} {72}
  (\bibinfo {year} {2023})},\ \bibinfo {note} {number: 1 Publisher: Nature
  Publishing Group}\BibitemShut {NoStop}%
\bibitem [{\citenamefont {Chang}\ \emph {et~al.}(2010)\citenamefont {Chang},
  \citenamefont {Betti}, \citenamefont {Spears}, \citenamefont {Anderson},
  \citenamefont {Edwards}, \citenamefont {Fatenejad}, \citenamefont {Lindl},
  \citenamefont {McCrory}, \citenamefont {Nora},\ and\ \citenamefont
  {Shvarts}}]{chang_generalized_2010}%
  \BibitemOpen
  \bibfield  {author} {\bibinfo {author} {\bibfnamefont {P.}~\bibnamefont
  {Chang}}, \bibinfo {author} {\bibfnamefont {R.}~\bibnamefont {Betti}},
  \bibinfo {author} {\bibfnamefont {B.~K.}\ \bibnamefont {Spears}}, \bibinfo
  {author} {\bibfnamefont {K.~S.}\ \bibnamefont {Anderson}}, \bibinfo {author}
  {\bibfnamefont {J.}~\bibnamefont {Edwards}}, \bibinfo {author} {\bibfnamefont
  {M.}~\bibnamefont {Fatenejad}}, \bibinfo {author} {\bibfnamefont {J.~D.}\
  \bibnamefont {Lindl}}, \bibinfo {author} {\bibfnamefont {R.~L.}\ \bibnamefont
  {McCrory}}, \bibinfo {author} {\bibfnamefont {R.}~\bibnamefont {Nora}}, \
  and\ \bibinfo {author} {\bibfnamefont {D.}~\bibnamefont {Shvarts}},\ }\href
  {\doibase 10.1103/PhysRevLett.104.135002} {\bibfield  {journal} {\bibinfo
  {journal} {Physical Review Letters}\ }\textbf {\bibinfo {volume} {104}},\
  \bibinfo {pages} {135002} (\bibinfo {year} {2010})}\BibitemShut {NoStop}%
\bibitem [{\citenamefont {Döppner}\ \emph {et~al.}(2015)\citenamefont
  {Döppner}, \citenamefont {Callahan}, \citenamefont {Hurricane},
  \citenamefont {Hinkel}, \citenamefont {Ma}, \citenamefont {Park},
  \citenamefont {Berzak~Hopkins}, \citenamefont {Casey}, \citenamefont
  {Celliers}, \citenamefont {Dewald}, \citenamefont {Dittrich}, \citenamefont
  {Haan}, \citenamefont {Kritcher}, \citenamefont {MacPhee}, \citenamefont
  {Le~Pape}, \citenamefont {Pak}, \citenamefont {Patel}, \citenamefont
  {Springer}, \citenamefont {Salmonson}, \citenamefont {Tommasini},
  \citenamefont {Benedetti}, \citenamefont {Bond}, \citenamefont {Bradley},
  \citenamefont {Caggiano}, \citenamefont {Church}, \citenamefont {Dixit},
  \citenamefont {Edgell}, \citenamefont {Edwards}, \citenamefont {Fittinghoff},
  \citenamefont {Frenje}, \citenamefont {Gatu~Johnson}, \citenamefont {Grim},
  \citenamefont {Hatarik}, \citenamefont {Havre}, \citenamefont {Herrmann},
  \citenamefont {Izumi}, \citenamefont {Khan}, \citenamefont {Kline},
  \citenamefont {Knauer}, \citenamefont {Kyrala}, \citenamefont {Landen},
  \citenamefont {Merrill}, \citenamefont {Moody}, \citenamefont {Moore},
  \citenamefont {Nikroo}, \citenamefont {Ralph}, \citenamefont {Remington},
  \citenamefont {Robey}, \citenamefont {Sayre}, \citenamefont {Schneider},
  \citenamefont {Streckert}, \citenamefont {Town}, \citenamefont {Turnbull},
  \citenamefont {Volegov}, \citenamefont {Wan}, \citenamefont {Widmann},
  \citenamefont {Wilde},\ and\ \citenamefont
  {Yeamans}}]{doppner_demonstration_2015}%
  \BibitemOpen
  \bibfield  {author} {\bibinfo {author} {\bibfnamefont {T.}~\bibnamefont
  {Döppner}}, \bibinfo {author} {\bibfnamefont {D.}~\bibnamefont {Callahan}},
  \bibinfo {author} {\bibfnamefont {O.}~\bibnamefont {Hurricane}}, \bibinfo
  {author} {\bibfnamefont {D.}~\bibnamefont {Hinkel}}, \bibinfo {author}
  {\bibfnamefont {T.}~\bibnamefont {Ma}}, \bibinfo {author} {\bibfnamefont
  {H.-S.}\ \bibnamefont {Park}}, \bibinfo {author} {\bibfnamefont
  {L.}~\bibnamefont {Berzak~Hopkins}}, \bibinfo {author} {\bibfnamefont
  {D.}~\bibnamefont {Casey}}, \bibinfo {author} {\bibfnamefont
  {P.}~\bibnamefont {Celliers}}, \bibinfo {author} {\bibfnamefont
  {E.}~\bibnamefont {Dewald}}, \bibinfo {author} {\bibfnamefont
  {T.}~\bibnamefont {Dittrich}}, \bibinfo {author} {\bibfnamefont
  {S.}~\bibnamefont {Haan}}, \bibinfo {author} {\bibfnamefont {A.}~\bibnamefont
  {Kritcher}}, \bibinfo {author} {\bibfnamefont {A.}~\bibnamefont {MacPhee}},
  \bibinfo {author} {\bibfnamefont {S.}~\bibnamefont {Le~Pape}}, \bibinfo
  {author} {\bibfnamefont {A.}~\bibnamefont {Pak}}, \bibinfo {author}
  {\bibfnamefont {P.}~\bibnamefont {Patel}}, \bibinfo {author} {\bibfnamefont
  {P.}~\bibnamefont {Springer}}, \bibinfo {author} {\bibfnamefont
  {J.}~\bibnamefont {Salmonson}}, \bibinfo {author} {\bibfnamefont
  {R.}~\bibnamefont {Tommasini}}, \bibinfo {author} {\bibfnamefont
  {L.}~\bibnamefont {Benedetti}}, \bibinfo {author} {\bibfnamefont
  {E.}~\bibnamefont {Bond}}, \bibinfo {author} {\bibfnamefont {D.}~\bibnamefont
  {Bradley}}, \bibinfo {author} {\bibfnamefont {J.}~\bibnamefont {Caggiano}},
  \bibinfo {author} {\bibfnamefont {J.}~\bibnamefont {Church}}, \bibinfo
  {author} {\bibfnamefont {S.}~\bibnamefont {Dixit}}, \bibinfo {author}
  {\bibfnamefont {D.}~\bibnamefont {Edgell}}, \bibinfo {author} {\bibfnamefont
  {M.}~\bibnamefont {Edwards}}, \bibinfo {author} {\bibfnamefont
  {D.}~\bibnamefont {Fittinghoff}}, \bibinfo {author} {\bibfnamefont
  {J.}~\bibnamefont {Frenje}}, \bibinfo {author} {\bibfnamefont
  {M.}~\bibnamefont {Gatu~Johnson}}, \bibinfo {author} {\bibfnamefont
  {G.}~\bibnamefont {Grim}}, \bibinfo {author} {\bibfnamefont {R.}~\bibnamefont
  {Hatarik}}, \bibinfo {author} {\bibfnamefont {M.}~\bibnamefont {Havre}},
  \bibinfo {author} {\bibfnamefont {H.}~\bibnamefont {Herrmann}}, \bibinfo
  {author} {\bibfnamefont {N.}~\bibnamefont {Izumi}}, \bibinfo {author}
  {\bibfnamefont {S.}~\bibnamefont {Khan}}, \bibinfo {author} {\bibfnamefont
  {J.}~\bibnamefont {Kline}}, \bibinfo {author} {\bibfnamefont
  {J.}~\bibnamefont {Knauer}}, \bibinfo {author} {\bibfnamefont
  {G.}~\bibnamefont {Kyrala}}, \bibinfo {author} {\bibfnamefont
  {O.}~\bibnamefont {Landen}}, \bibinfo {author} {\bibfnamefont
  {F.}~\bibnamefont {Merrill}}, \bibinfo {author} {\bibfnamefont
  {J.}~\bibnamefont {Moody}}, \bibinfo {author} {\bibfnamefont
  {A.}~\bibnamefont {Moore}}, \bibinfo {author} {\bibfnamefont
  {A.}~\bibnamefont {Nikroo}}, \bibinfo {author} {\bibfnamefont
  {J.}~\bibnamefont {Ralph}}, \bibinfo {author} {\bibfnamefont
  {B.}~\bibnamefont {Remington}}, \bibinfo {author} {\bibfnamefont
  {H.}~\bibnamefont {Robey}}, \bibinfo {author} {\bibfnamefont
  {D.}~\bibnamefont {Sayre}}, \bibinfo {author} {\bibfnamefont
  {M.}~\bibnamefont {Schneider}}, \bibinfo {author} {\bibfnamefont
  {H.}~\bibnamefont {Streckert}}, \bibinfo {author} {\bibfnamefont
  {R.}~\bibnamefont {Town}}, \bibinfo {author} {\bibfnamefont {D.}~\bibnamefont
  {Turnbull}}, \bibinfo {author} {\bibfnamefont {P.}~\bibnamefont {Volegov}},
  \bibinfo {author} {\bibfnamefont {A.}~\bibnamefont {Wan}}, \bibinfo {author}
  {\bibfnamefont {K.}~\bibnamefont {Widmann}}, \bibinfo {author} {\bibfnamefont
  {C.}~\bibnamefont {Wilde}}, \ and\ \bibinfo {author} {\bibfnamefont
  {C.}~\bibnamefont {Yeamans}},\ }\href {\doibase
  10.1103/PhysRevLett.115.055001} {\bibfield  {journal} {\bibinfo  {journal}
  {Physical Review Letters}\ }\textbf {\bibinfo {volume} {115}},\ \bibinfo
  {pages} {055001} (\bibinfo {year} {2015})}\BibitemShut {NoStop}%
\bibitem [{\citenamefont {Gopalaswamy}\ \emph {et~al.}(2024)\citenamefont
  {Gopalaswamy}, \citenamefont {Williams}, \citenamefont {Betti}, \citenamefont
  {Patel}, \citenamefont {Knauer}, \citenamefont {Lees}, \citenamefont {Cao},
  \citenamefont {Campbell}, \citenamefont {Farmakis}, \citenamefont {Ejaz},
  \citenamefont {Anderson}, \citenamefont {Epstein}, \citenamefont
  {Carroll-Nellenbeck}, \citenamefont {Igumenshchev}, \citenamefont {Marozas},
  \citenamefont {Radha}, \citenamefont {Solodov}, \citenamefont {Thomas},
  \citenamefont {Woo}, \citenamefont {Collins}, \citenamefont {Hu},
  \citenamefont {Scullin}, \citenamefont {Turnbull}, \citenamefont {Goncharov},
  \citenamefont {Churnetski}, \citenamefont {Forrest}, \citenamefont {Glebov},
  \citenamefont {Heuer}, \citenamefont {McClow}, \citenamefont {Shah},
  \citenamefont {Stoeckl}, \citenamefont {Theobald}, \citenamefont {Edgell},
  \citenamefont {Ivancic}, \citenamefont {Rosenberg}, \citenamefont {Regan},
  \citenamefont {Bredesen}, \citenamefont {Fella}, \citenamefont {Koch},
  \citenamefont {Janezic}, \citenamefont {Bonino}, \citenamefont {Harding},
  \citenamefont {Bauer}, \citenamefont {Sampat}, \citenamefont {Waxer},
  \citenamefont {Labuzeta}, \citenamefont {Morse}, \citenamefont
  {Gatu-Johnson}, \citenamefont {Petrasso}, \citenamefont {Frenje},
  \citenamefont {Murray}, \citenamefont {Serrato}, \citenamefont {Guzman},
  \citenamefont {Shuldberg}, \citenamefont {Farrell},\ and\ \citenamefont
  {Deeney}}]{gopalaswamy_demonstration_2024}%
  \BibitemOpen
  \bibfield  {author} {\bibinfo {author} {\bibfnamefont {V.}~\bibnamefont
  {Gopalaswamy}}, \bibinfo {author} {\bibfnamefont {C.~A.}\ \bibnamefont
  {Williams}}, \bibinfo {author} {\bibfnamefont {R.}~\bibnamefont {Betti}},
  \bibinfo {author} {\bibfnamefont {D.}~\bibnamefont {Patel}}, \bibinfo
  {author} {\bibfnamefont {J.~P.}\ \bibnamefont {Knauer}}, \bibinfo {author}
  {\bibfnamefont {A.}~\bibnamefont {Lees}}, \bibinfo {author} {\bibfnamefont
  {D.}~\bibnamefont {Cao}}, \bibinfo {author} {\bibfnamefont {E.~M.}\
  \bibnamefont {Campbell}}, \bibinfo {author} {\bibfnamefont {P.}~\bibnamefont
  {Farmakis}}, \bibinfo {author} {\bibfnamefont {R.}~\bibnamefont {Ejaz}},
  \bibinfo {author} {\bibfnamefont {K.~S.}\ \bibnamefont {Anderson}}, \bibinfo
  {author} {\bibfnamefont {R.}~\bibnamefont {Epstein}}, \bibinfo {author}
  {\bibfnamefont {J.}~\bibnamefont {Carroll-Nellenbeck}}, \bibinfo {author}
  {\bibfnamefont {I.~V.}\ \bibnamefont {Igumenshchev}}, \bibinfo {author}
  {\bibfnamefont {J.~A.}\ \bibnamefont {Marozas}}, \bibinfo {author}
  {\bibfnamefont {P.~B.}\ \bibnamefont {Radha}}, \bibinfo {author}
  {\bibfnamefont {A.~A.}\ \bibnamefont {Solodov}}, \bibinfo {author}
  {\bibfnamefont {C.~A.}\ \bibnamefont {Thomas}}, \bibinfo {author}
  {\bibfnamefont {K.~M.}\ \bibnamefont {Woo}}, \bibinfo {author} {\bibfnamefont
  {T.~J.~B.}\ \bibnamefont {Collins}}, \bibinfo {author} {\bibfnamefont
  {S.~X.}\ \bibnamefont {Hu}}, \bibinfo {author} {\bibfnamefont
  {W.}~\bibnamefont {Scullin}}, \bibinfo {author} {\bibfnamefont
  {D.}~\bibnamefont {Turnbull}}, \bibinfo {author} {\bibfnamefont {V.~N.}\
  \bibnamefont {Goncharov}}, \bibinfo {author} {\bibfnamefont {K.}~\bibnamefont
  {Churnetski}}, \bibinfo {author} {\bibfnamefont {C.~J.}\ \bibnamefont
  {Forrest}}, \bibinfo {author} {\bibfnamefont {V.~Y.}\ \bibnamefont {Glebov}},
  \bibinfo {author} {\bibfnamefont {P.~V.}\ \bibnamefont {Heuer}}, \bibinfo
  {author} {\bibfnamefont {H.}~\bibnamefont {McClow}}, \bibinfo {author}
  {\bibfnamefont {R.~C.}\ \bibnamefont {Shah}}, \bibinfo {author}
  {\bibfnamefont {C.}~\bibnamefont {Stoeckl}}, \bibinfo {author} {\bibfnamefont
  {W.}~\bibnamefont {Theobald}}, \bibinfo {author} {\bibfnamefont {D.~H.}\
  \bibnamefont {Edgell}}, \bibinfo {author} {\bibfnamefont {S.}~\bibnamefont
  {Ivancic}}, \bibinfo {author} {\bibfnamefont {M.~J.}\ \bibnamefont
  {Rosenberg}}, \bibinfo {author} {\bibfnamefont {S.~P.}\ \bibnamefont
  {Regan}}, \bibinfo {author} {\bibfnamefont {D.}~\bibnamefont {Bredesen}},
  \bibinfo {author} {\bibfnamefont {C.}~\bibnamefont {Fella}}, \bibinfo
  {author} {\bibfnamefont {M.}~\bibnamefont {Koch}}, \bibinfo {author}
  {\bibfnamefont {R.~T.}\ \bibnamefont {Janezic}}, \bibinfo {author}
  {\bibfnamefont {M.~J.}\ \bibnamefont {Bonino}}, \bibinfo {author}
  {\bibfnamefont {D.~R.}\ \bibnamefont {Harding}}, \bibinfo {author}
  {\bibfnamefont {K.~A.}\ \bibnamefont {Bauer}}, \bibinfo {author}
  {\bibfnamefont {S.}~\bibnamefont {Sampat}}, \bibinfo {author} {\bibfnamefont
  {L.~J.}\ \bibnamefont {Waxer}}, \bibinfo {author} {\bibfnamefont
  {M.}~\bibnamefont {Labuzeta}}, \bibinfo {author} {\bibfnamefont {S.~F.~B.}\
  \bibnamefont {Morse}}, \bibinfo {author} {\bibfnamefont {M.}~\bibnamefont
  {Gatu-Johnson}}, \bibinfo {author} {\bibfnamefont {R.~D.}\ \bibnamefont
  {Petrasso}}, \bibinfo {author} {\bibfnamefont {J.~A.}\ \bibnamefont
  {Frenje}}, \bibinfo {author} {\bibfnamefont {J.}~\bibnamefont {Murray}},
  \bibinfo {author} {\bibfnamefont {B.}~\bibnamefont {Serrato}}, \bibinfo
  {author} {\bibfnamefont {D.}~\bibnamefont {Guzman}}, \bibinfo {author}
  {\bibfnamefont {C.}~\bibnamefont {Shuldberg}}, \bibinfo {author}
  {\bibfnamefont {M.}~\bibnamefont {Farrell}}, \ and\ \bibinfo {author}
  {\bibfnamefont {C.}~\bibnamefont {Deeney}},\ }\href {\doibase
  10.1038/s41567-023-02361-4} {\bibfield  {journal} {\bibinfo  {journal}
  {Nature Physics}\ ,\ \bibinfo {pages} {1}} (\bibinfo {year} {2024})},\
  \bibinfo {note} {publisher: Nature Publishing Group}\BibitemShut {NoStop}%
\bibitem [{\citenamefont {Churnetski}\ \emph {et~al.}(2024)\citenamefont
  {Churnetski}, \citenamefont {Woo}, \citenamefont {Theobald}, \citenamefont
  {Stoeckl}, \citenamefont {Ceurvorst}, \citenamefont {Gopalaswamy},
  \citenamefont {Rinderknecht}, \citenamefont {Heuer}, \citenamefont {Knauer},
  \citenamefont {Forrest}, \citenamefont {Igumenshchev}, \citenamefont
  {Ivancic}, \citenamefont {Michalko}, \citenamefont {Shah}, \citenamefont
  {Lees}, \citenamefont {Bahukutumbi}, \citenamefont {Betti}, \citenamefont
  {Thomas}, \citenamefont {Regan}, \citenamefont {Kunimune}, \citenamefont
  {Wink}, \citenamefont {Adrian}, \citenamefont {Gatu~Johnson},\ and\
  \citenamefont {Frenje}}]{churnetski_three-dimensional_2024}%
  \BibitemOpen
  \bibfield  {author} {\bibinfo {author} {\bibfnamefont {K.}~\bibnamefont
  {Churnetski}}, \bibinfo {author} {\bibfnamefont {K.~M.}\ \bibnamefont {Woo}},
  \bibinfo {author} {\bibfnamefont {W.}~\bibnamefont {Theobald}}, \bibinfo
  {author} {\bibfnamefont {C.}~\bibnamefont {Stoeckl}}, \bibinfo {author}
  {\bibfnamefont {L.}~\bibnamefont {Ceurvorst}}, \bibinfo {author}
  {\bibfnamefont {V.}~\bibnamefont {Gopalaswamy}}, \bibinfo {author}
  {\bibfnamefont {H.}~\bibnamefont {Rinderknecht}}, \bibinfo {author}
  {\bibfnamefont {P.~V.}\ \bibnamefont {Heuer}}, \bibinfo {author}
  {\bibfnamefont {J.}~\bibnamefont {Knauer}}, \bibinfo {author} {\bibfnamefont
  {C.}~\bibnamefont {Forrest}}, \bibinfo {author} {\bibfnamefont
  {I.}~\bibnamefont {Igumenshchev}}, \bibinfo {author} {\bibfnamefont
  {S.}~\bibnamefont {Ivancic}}, \bibinfo {author} {\bibfnamefont
  {M.}~\bibnamefont {Michalko}}, \bibinfo {author} {\bibfnamefont
  {R.}~\bibnamefont {Shah}}, \bibinfo {author} {\bibfnamefont {A.}~\bibnamefont
  {Lees}}, \bibinfo {author} {\bibfnamefont {R.}~\bibnamefont {Bahukutumbi}},
  \bibinfo {author} {\bibfnamefont {R.}~\bibnamefont {Betti}}, \bibinfo
  {author} {\bibfnamefont {C.}~\bibnamefont {Thomas}}, \bibinfo {author}
  {\bibfnamefont {S.}~\bibnamefont {Regan}}, \bibinfo {author} {\bibfnamefont
  {J.}~\bibnamefont {Kunimune}}, \bibinfo {author} {\bibfnamefont
  {C.}~\bibnamefont {Wink}}, \bibinfo {author} {\bibfnamefont {P.}~\bibnamefont
  {Adrian}}, \bibinfo {author} {\bibfnamefont {M.}~\bibnamefont
  {Gatu~Johnson}}, \ and\ \bibinfo {author} {\bibfnamefont {J.}~\bibnamefont
  {Frenje}},\ }\href {\doibase 10.2139/ssrn.4740997} {\emph {\bibinfo {title}
  {Three-{Dimensional} {Reconstruction} of {Implosion} {Stagnation} in {Laser}
  {Direct} {Drive} on {Omega}}}},\ \bibinfo {type} {preprint}\ (\bibinfo
  {institution} {SSRN},\ \bibinfo {year} {2024})\BibitemShut {NoStop}%
\bibitem [{\citenamefont {Atzeni}\ and\ \citenamefont {Meyer-ter
  Vehn}()}]{atzeni_physics_2004}%
  \BibitemOpen
  \bibfield  {author} {\bibinfo {author} {\bibfnamefont {S.}~\bibnamefont
  {Atzeni}}\ and\ \bibinfo {author} {\bibfnamefont {J.}~\bibnamefont {Meyer-ter
  Vehn}},\ }\href@noop {} {\emph {\bibinfo {title} {The Physics of Inertial
  Fusion: Beam Plasma Interaction, Hydrodynamics, Hot Dense Matter}}},\
  \bibinfo {series} {Oxford Science Publications}\ No.\ \bibinfo {number}
  {125}\ (\bibinfo  {publisher} {Clarendon Press ; Oxford University
  Press})\BibitemShut {NoStop}%
\bibitem [{\citenamefont {Hurricane}\ \emph {et~al.}(2023)\citenamefont
  {Hurricane}, \citenamefont {Patel}, \citenamefont {Betti}, \citenamefont
  {Froula}, \citenamefont {Regan}, \citenamefont {Slutz}, \citenamefont
  {Gomez},\ and\ \citenamefont {Sweeney}}]{hurricane_physics_2023}%
  \BibitemOpen
  \bibfield  {author} {\bibinfo {author} {\bibfnamefont {O.}~\bibnamefont
  {Hurricane}}, \bibinfo {author} {\bibfnamefont {P.}~\bibnamefont {Patel}},
  \bibinfo {author} {\bibfnamefont {R.}~\bibnamefont {Betti}}, \bibinfo
  {author} {\bibfnamefont {D.}~\bibnamefont {Froula}}, \bibinfo {author}
  {\bibfnamefont {S.}~\bibnamefont {Regan}}, \bibinfo {author} {\bibfnamefont
  {S.}~\bibnamefont {Slutz}}, \bibinfo {author} {\bibfnamefont
  {M.}~\bibnamefont {Gomez}}, \ and\ \bibinfo {author} {\bibfnamefont
  {M.}~\bibnamefont {Sweeney}},\ }\href {\doibase 10.1103/RevModPhys.95.025005}
  {\bibfield  {journal} {\bibinfo  {journal} {Reviews of Modern Physics}\
  }\textbf {\bibinfo {volume} {95}},\ \bibinfo {pages} {025005} (\bibinfo
  {year} {2023})}\BibitemShut {NoStop}%
\bibitem [{\citenamefont {Daughton}\ \emph {et~al.}(2023)\citenamefont
  {Daughton}, \citenamefont {Albright}, \citenamefont {Finnegan}, \citenamefont
  {Haines}, \citenamefont {Kline}, \citenamefont {Sauppe},\ and\ \citenamefont
  {Smidt}}]{daughton_infuence_2023}%
  \BibitemOpen
  \bibfield  {author} {\bibinfo {author} {\bibfnamefont {W.}~\bibnamefont
  {Daughton}}, \bibinfo {author} {\bibfnamefont {B.~J.}\ \bibnamefont
  {Albright}}, \bibinfo {author} {\bibfnamefont {S.~M.}\ \bibnamefont
  {Finnegan}}, \bibinfo {author} {\bibfnamefont {B.~M.}\ \bibnamefont
  {Haines}}, \bibinfo {author} {\bibfnamefont {J.~L.}\ \bibnamefont {Kline}},
  \bibinfo {author} {\bibfnamefont {J.~P.}\ \bibnamefont {Sauppe}}, \ and\
  \bibinfo {author} {\bibfnamefont {J.~M.}\ \bibnamefont {Smidt}},\ }\href
  {\doibase 10.1063/5.0129561} {\bibfield  {journal} {\bibinfo  {journal}
  {Physics of Plasmas}\ }\textbf {\bibinfo {volume} {30}},\ \bibinfo {pages}
  {012704} (\bibinfo {year} {2023})},\ \bibinfo {note} {arXiv:2207.00093
  [physics]}\BibitemShut {NoStop}%
\bibitem [{\citenamefont {Fan}\ \emph {et~al.}(2016)\citenamefont {Fan},
  \citenamefont {Liu}, \citenamefont {Liu}, \citenamefont {Yu},\ and\
  \citenamefont {He}}]{fan_ignition_2016}%
  \BibitemOpen
  \bibfield  {author} {\bibinfo {author} {\bibfnamefont {Z.}~\bibnamefont
  {Fan}}, \bibinfo {author} {\bibfnamefont {J.}~\bibnamefont {Liu}}, \bibinfo
  {author} {\bibfnamefont {B.}~\bibnamefont {Liu}}, \bibinfo {author}
  {\bibfnamefont {C.}~\bibnamefont {Yu}}, \ and\ \bibinfo {author}
  {\bibfnamefont {X.~T.}\ \bibnamefont {He}},\ }\href {\doibase
  10.1063/1.4940315} {\bibfield  {journal} {\bibinfo  {journal} {Physics of
  Plasmas}\ }\textbf {\bibinfo {volume} {23}},\ \bibinfo {pages} {010703}
  (\bibinfo {year} {2016})}\BibitemShut {NoStop}%
\bibitem [{\citenamefont {Fraley}\ \emph {et~al.}(1974)\citenamefont {Fraley},
  \citenamefont {Linnebur}, \citenamefont {Mason},\ and\ \citenamefont
  {Morse}}]{fraley_thermonuclear_1974}%
  \BibitemOpen
  \bibfield  {author} {\bibinfo {author} {\bibfnamefont {G.~S.}\ \bibnamefont
  {Fraley}}, \bibinfo {author} {\bibfnamefont {E.~J.}\ \bibnamefont
  {Linnebur}}, \bibinfo {author} {\bibfnamefont {R.~J.}\ \bibnamefont {Mason}},
  \ and\ \bibinfo {author} {\bibfnamefont {R.~L.}\ \bibnamefont {Morse}},\
  }\href {\doibase 10.1063/1.1694739} {\bibfield  {journal} {\bibinfo
  {journal} {The Physics of Fluids}\ }\textbf {\bibinfo {volume} {17}},\
  \bibinfo {pages} {474} (\bibinfo {year} {1974})}\BibitemShut {NoStop}%
\bibitem [{\citenamefont {Wu}\ \emph {et~al.}(2018)\citenamefont {Wu},
  \citenamefont {He}, \citenamefont {Yu},\ and\ \citenamefont
  {Fritzsche}}]{wu_particle--cell_2018}%
  \BibitemOpen
  \bibfield  {author} {\bibinfo {author} {\bibfnamefont {D.}~\bibnamefont
  {Wu}}, \bibinfo {author} {\bibfnamefont {X.~T.}\ \bibnamefont {He}}, \bibinfo
  {author} {\bibfnamefont {W.}~\bibnamefont {Yu}}, \ and\ \bibinfo {author}
  {\bibfnamefont {S.}~\bibnamefont {Fritzsche}},\ }\href {\doibase
  10.1017/hpl.2018.41} {\bibfield  {journal} {\bibinfo  {journal} {High Power
  Laser Science and Engineering}\ }\textbf {\bibinfo {volume} {6}},\ \bibinfo
  {pages} {e50} (\bibinfo {year} {2018})}\BibitemShut {NoStop}%
\bibitem [{\citenamefont {Spears}\ \emph {et~al.}(2008)\citenamefont {Spears},
  \citenamefont {Hicks}, \citenamefont {Velsko}, \citenamefont {Stoyer},
  \citenamefont {Robey}, \citenamefont {Munro}, \citenamefont {Haan},
  \citenamefont {Landen}, \citenamefont {Nikroo},\ and\ \citenamefont
  {Huang}}]{spears_influence_2008}%
  \BibitemOpen
  \bibfield  {author} {\bibinfo {author} {\bibfnamefont {B.}~\bibnamefont
  {Spears}}, \bibinfo {author} {\bibfnamefont {D.}~\bibnamefont {Hicks}},
  \bibinfo {author} {\bibfnamefont {C.}~\bibnamefont {Velsko}}, \bibinfo
  {author} {\bibfnamefont {M.}~\bibnamefont {Stoyer}}, \bibinfo {author}
  {\bibfnamefont {H.}~\bibnamefont {Robey}}, \bibinfo {author} {\bibfnamefont
  {D.}~\bibnamefont {Munro}}, \bibinfo {author} {\bibfnamefont
  {S.}~\bibnamefont {Haan}}, \bibinfo {author} {\bibfnamefont {O.}~\bibnamefont
  {Landen}}, \bibinfo {author} {\bibfnamefont {A.}~\bibnamefont {Nikroo}}, \
  and\ \bibinfo {author} {\bibfnamefont {H.}~\bibnamefont {Huang}},\ }\href
  {\doibase 10.1088/1742-6596/112/2/022003} {\bibfield  {journal} {\bibinfo
  {journal} {Journal of Physics: Conference Series}\ }\textbf {\bibinfo
  {volume} {112}},\ \bibinfo {pages} {022003} (\bibinfo {year}
  {2008})}\BibitemShut {NoStop}%
\bibitem [{\citenamefont {Atzeni}\ and\ \citenamefont
  {Caruso}()}]{atzeni_inertial_nodate}%
  \BibitemOpen
  \bibfield  {author} {\bibinfo {author} {\bibfnamefont {S.}~\bibnamefont
  {Atzeni}}\ and\ \bibinfo {author} {\bibfnamefont {A.}~\bibnamefont
  {Caruso}},\ }\href@noop {} {\ }\BibitemShut {NoStop}%
\bibitem [{\citenamefont {Daligault}\ and\ \citenamefont
  {Simoni}(2019)}]{daligault_theory_2019}%
  \BibitemOpen
  \bibfield  {author} {\bibinfo {author} {\bibfnamefont {J.}~\bibnamefont
  {Daligault}}\ and\ \bibinfo {author} {\bibfnamefont {J.}~\bibnamefont
  {Simoni}},\ }\href {\doibase 10.1103/PhysRevE.100.043201} {\bibfield
  {journal} {\bibinfo  {journal} {Physical Review E}\ }\textbf {\bibinfo
  {volume} {100}},\ \bibinfo {pages} {043201} (\bibinfo {year} {2019})},\
  \bibinfo {note} {publisher: American Physical Society}\BibitemShut {NoStop}%
\bibitem [{\citenamefont {Temporal}\ \emph {et~al.}(2012)\citenamefont
  {Temporal}, \citenamefont {Brandon}, \citenamefont {Canaud}, \citenamefont
  {Didelez}, \citenamefont {Fedosejevs},\ and\ \citenamefont
  {Ramis}}]{temporal_ignition_2012}%
  \BibitemOpen
  \bibfield  {author} {\bibinfo {author} {\bibfnamefont {M.}~\bibnamefont
  {Temporal}}, \bibinfo {author} {\bibfnamefont {V.}~\bibnamefont {Brandon}},
  \bibinfo {author} {\bibfnamefont {B.}~\bibnamefont {Canaud}}, \bibinfo
  {author} {\bibfnamefont {J.}~\bibnamefont {Didelez}}, \bibinfo {author}
  {\bibfnamefont {R.}~\bibnamefont {Fedosejevs}}, \ and\ \bibinfo {author}
  {\bibfnamefont {R.}~\bibnamefont {Ramis}},\ }\href {\doibase
  10.1088/0029-5515/52/10/103011} {\bibfield  {journal} {\bibinfo  {journal}
  {Nuclear Fusion}\ }\textbf {\bibinfo {volume} {52}},\ \bibinfo {pages}
  {103011} (\bibinfo {year} {2012})}\BibitemShut {NoStop}%
\end{thebibliography}%
% \end{thebibliography}

\end{document}